\documentclass{pasj00}

\begin{document}

\SetRunningHead{Shimasaku et al.}{Lyman $\alpha$ emitters at $z=5.7$}
\Received{2005/10/26}
\Accepted{2006/02/27}

\title{Ly$\alpha$ Emitters at $z=5.7$ in the Subaru Deep Field
\altaffilmark{*,\dag}
}

\author{%
  Kazuhiro  \textsc{Shimasaku}, \altaffilmark{1,2}
  Nobunari  \textsc{Kashikawa}, \altaffilmark{3,4} 
  Mamoru    \textsc{Doi},       \altaffilmark{5,2} 
  Chun      \textsc{Ly},        \altaffilmark{6} \\ 
  Matthew A.\textsc{Malkan},    \altaffilmark{6} 
  Yuichi    \textsc{Matsuda},   \altaffilmark{7} 
  Masami    \textsc{Ouchi},     \altaffilmark{8}\altaffilmark{,\ddag} \\
  Tomoki    \textsc{Hayashino}, \altaffilmark{9}
  Masanori  \textsc{Iye},       \altaffilmark{3,4}
  Kentaro   \textsc{Motohara},  \altaffilmark{5}
  Takashi   \textsc{Murayama},  \altaffilmark{10} \\
  Tohru     \textsc{Nagao},     \altaffilmark{3,11} 
  Kouji     \textsc{Ohta},      \altaffilmark{12} 
  Sadanori  \textsc{Okamura},   \altaffilmark{1,2}
  Toshiyuki \textsc{Sasaki},    \altaffilmark{13} \\
  Yasuhiro  \textsc{Shioya},    \altaffilmark{10} 
  Yoshiaki  \textsc{Taniguchi}  \altaffilmark{10}
}
\altaffiltext{1}{Department of Astronomy, School of Science,
        The University of Tokyo, Tokyo 113-0033
        \\ Email (KS): shimasaku@astron.s.u-tokyo.ac.jp}
\altaffiltext{2}{Research Center for the Early Universe, 
        School of Science,
        The University of Tokyo, \\ Tokyo 113-0033}
\altaffiltext{3}{Optical and Infrared Astronomy Division, 
        National Astronomical Observatory, 
        Mitaka, \\ Tokyo 181-8588}
\altaffiltext{4}{Department of Astronomy, School of Science,
        Graduate University for Advanced Studies, \\
        Mitaka, Tokyo 181-8588}
\altaffiltext{5}{Institute of Astronomy, School of Science, 
        The University of Tokyo, Mitaka 181-0015}
\altaffiltext{6}{Department of Astronomy, 
        University of California at Los Angeles, 
        Los Angeles, \\ CA 90095-1547, USA}
\altaffiltext{7}{Department of Astronomy, 
        Graduate School of Science, Kyoto University, 
        Kyoto 606-8502}
\altaffiltext{8}{Space Telescope Science Institute, 
        3700 San Martin Drive,
        Baltimore, MD 21218, USA}
\altaffiltext{9}{Research Center for Neutrino Science,
        Graduate School of Science, Tohoku University, \\
        Aramaki, Aoba, Sendai 980-8578}
\altaffiltext{10}{Astronomical Institute, Graduate School of Science,
        Tohoku University, Aramaki, Aoba, \\ Sendai 980-8578, Japan}
\altaffiltext{11}{INAF / Osservatorio Astrofisico di Arcetri
        Largo E. Fermi 5, 50125 Firenze, Italy}
\altaffiltext{12}{Kouji Ohta Department of Astronomy, 
        Kyoto University, Kyoto 606-8502}
\altaffiltext{13}{Subaru Telescope, 
        National Astronomical Observatory of Japan, 
        650 N. A'ohoku Place, \\ 
        Hilo, HI 96720, USA}
%
%

\maketitle

\begin{abstract}
We present the properties of Ly$\alpha$ emitters (LAEs) at $z=5.7$ 
in the Subaru Deep Field. 
A photometric sample of 89 LAE candidates is constructed from 
narrow-band (NB816) data down to $NB816 = 26.0$ (AB) 
in a continuous 725 arcmin$^2$ area.
Spectra of 39 objects satisfying the photometric selection 
criteria for LAEs were obtained with Subaru and Keck II Telescopes, 
among which 28 were confirmed LAEs, 
one was a nearby galaxy, and eight were unclassified. 
We also obtained spectra of another 24 NB816-excess objects 
in the field, identifying six additional LAEs.
We find that the Ly$\alpha$ luminosity function derived from 
the photometric sample 
is reproduced well by a Schechter function with 
$L^\star = 7.9^{+3.0}_{-2.2} \times 10^{42}$ erg s$^{-1}$ 
and $\phi^\star = 6.3^{+3.0}_{-2.0} \times 10^{-4}$ Mpc$^{-3}$ 
for $\alpha = -1.5$ (fixed) 
over the whole luminosity range of 
$L \simeq 3 \times 10^{42}$ -- $3 \times 10^{43}$ erg s$^{-1}$.
We then measure rest-frame Ly$\alpha$ equivalent widths (EWs) 
for the confirmed LAEs, 
to find that the median among the 28 objects satisfying 
the photometric selection criteria is $W{\rm_{0}^{i}} = 233$ \AA. 
We infer that $30\%$ -- $40\%$ of LAEs at $z=5.7$ 
exceed $W{\rm_{0}^{i}} = 240$ \AA.
These large-EW objects probably cannot be accounted for by 
ordinary star-forming populations with a Salpeter IMF.
We also find that LAEs with fainter far-UV luminosities 
have larger EWs.
Finally, we derive the far-UV luminosity function 
of LAEs down to $M_{\rm UV} \simeq -19.6$ using the photometric 
sample, and compare it with that of Lyman-break galaxies (LBGs).
We find that as high as about $80\%$ of LBGs at $z\sim 6$ have 
$W{\rm_{0}^{i}} \ge 100$ \AA, 
in sharp contrast to lower-$z$ counterparts.

\end{abstract}

\clearpage 

\footnotetext[*]
{Based on data collected at the Subaru Telescope, which is 
operated by the National \\ 
Astronomical Observatory of Japan.}

\footnotetext[\dag]
{Part of the data presented herein were obtained at the W.M. Keck 
Observatory, which is \\ 
operated as a scientific partnership among the California Institute 
of Technology, \\ 
the University of California and the National Aeronautics and Space 
Administration. \\
The Observatory was made possible by the generous financial support \\
of the W.M. Keck Foundation.}

\footnotetext[\ddag]
{Hubble Fellow.}

%
%

\section{Introduction}

Ly$\alpha$ emitters (LAEs) are commonly seen 
in the high-redshift universe.
Most of them are thought to be very young star-forming galaxies, 
although some may have an active nucleus 
and some could be primeval galaxies in a gas cooling phase 
before onset of the initial star formation or in superwind 
phases just after the initial star formation. 
The high number density of LAEs, roughly comparable 
to that of Lyman-break galaxies (LBGs), suggests that they 
are important in understanding the early evolution 
of galaxies.

LAEs at $z>5$ are of particular interest in galaxy formation 
and cosmology, since they are not only expected to be extremely 
young but also an invaluable probe of the universe during or 
just after reionization.
They can be used to infer the reionization epoch 
of the universe via the change in their number density 
(Haiman \& Spaans 1999; Rhoads \& Malhotra 2001; 
Haiman 2002; Malhotra \& Rhoads 2004; Stern et al. 2005; 
Kashikawa et al. 2005).
However, although there has been rapid progress 
in the observation of LAEs at $z>5$ 
(e.g., Rhoads \& Malhotra 2001; 
Ajiki et al. 2003, 2004; Kodaira et al. 2003; Hu et al. 2002, 2004; 
Maier et al. 2003; Rhoads et al. 2003; Ouchi et al. 2005; 
Taniguchi et al. 2005), 
the data obtained to date are not sufficient 
to place strong constraints on their fundamental properties 
such as the luminosity function (LF), 
in contrast to the data of LBGs at similar redshifts.
Usually, narrow-band surveys are used to detect LAEs.
Their apparent faintness in both narrow bands (i.e., Ly$\alpha$) 
and broad bands (off Ly$\alpha$), 
as well as a small survey volume defined by the wavelength coverage 
of the narrow-band filter times the survey area, 
makes it difficult to obtain a large sample. 
Samples from several small surveys by different workers 
could be combined to form a larger sample.
However, since the selection criteria of LAEs and the survey 
depth differ among the surveys, LFs calculated from such combined 
samples may have systematic uncertainties difficult to control.

In this paper, we report on the results of a very deep survey 
of LAEs at $z=5.7$ in a wide, continuous area of 725 arcmin$^2$.
The survey uses deep imaging data in six optical bandpasses   
including the narrow band NB816 ($\lambda_c=8150$ \AA\ 
and FWHM $=120$ \AA) targeted at the Ly$\alpha$ emission 
at $z \simeq 5.7 \pm 0.05$. 
All of the imaging data were taken with the Suprime-Cam 
(Miyazaki et al. 2002) at the Subaru Telescope (Iye et al. 2004).
From these data, we construct a sample of 89 photometrically 
selected LAEs down to $NB816 = 26.0$.
By followup spectroscopy with the Subaru and Keck II Telescopes, 
we obtained a sample of 34 LAEs with confirmed redshifts. 

Combining these samples with Monte Carlo simulations, 
we calculate the Ly$\alpha$ LF, 
which is the most reliable measurement for the LF of 
$z=5.7$ LAEs published to date.
This LF serves as a {\lq}zero point{\rq} for the LF 
in the fully ionized universe ($z<6$), which is to be compared 
with the LF at $z>6$ to probe the reionization history 
of the universe.
A companion paper (Kashikawa et al. 2005)  
compares the LF at $z=5.7$ obtained in this paper with that 
for $z=6.5$ LAEs, and discusses the ionization state of 
the intergalactic medium at $z=6.5$.

We also derive accurately the far UV continuum LF 
and the distribution of Ly$\alpha$ equivalent widths (EWs), 
both of which have not yet been strongly constrained 
for LAEs at $z>5$.
Comparing the far UV LF of our LAEs with that of LBGs at $z \sim 6$ 
given in the literature, 
we examine what fraction of LBGs have EWs large enough to be 
detected as LAEs, and discuss evolution of Ly$\alpha$ properties 
of star-forming galaxies.

The plan of this paper is as follows.
Section 2 describes the photometric and spectroscopic data 
used in this study.
The confirmed sample of 34 LAEs is constructed in \S 3.
The photometric selection of LAEs is made in \S 4.
Section 5 is devoted to results and discussion, 
and a summary is given in \S 6.
AB magnitudes are used in this paper.
We assume a flat universe with 
$\Omega_{\rm M}=0.3$, $\Omega_\Lambda=0.7$, 
and $H_0 = 70$ km s$^{-1}$ Mpc$^{-1}$.

%
%

\section{Data}

\subsection{Imaging Data}

The SDF has deep, public Subaru/Suprime-Cam data in seven bandpasses, 
$B$, $V$, $R$, $i'$, $z'$, NB816, and NB921, 
obtained for the Subaru Deep Field Project 
(Kashikawa et al. 2004; 
see also Maihara et al. 2001 for the Subaru Deep Field).
This project is a large program of Subaru Observatory to carry out 
a deep galaxy survey in a blank field in optical and NIR 
wavelengths to study distant galaxies.
$B$, $V$, $R$, $i'$, and $z'$ are standard Johnson and SDSS 
broad bands, and NB816 and NB921 are narrow bands 
whose central wavelength and FWHM are 
(8150 \AA, 120 \AA) and (9196 \AA, 132 \AA), respectively.
In this paper, we use the $B$, $V$, $R$, $i'$, $z'$, and NB816 data 
to study LAEs at $z=5.7$.
The exposure times and the $3\sigma$ limiting magnitudes 
on a $2''$ aperture are: 
595 min and 28.45 mag ($B$),
340 min and 27.74 mag ($V$),
600 min and 27.80 mag ($R$),
801 min and 27.43 mag ($i'$),
504 min and 26.62 mag ($z'$),
600 min and 26.63 mag (NB816).
These images have been convolved to a seeing of 
$0.''98$ (FWHM) and an identical sky coverage of $29.'7 \times 36.'7$.
The pixel scale of the images is $0.''202$ pixel$^{-1}$.
The selection of $z=5.7$ LAEs is made 
in the $R-z'$ vs $i'-NB816$ plane, 
and the $B$ and $V$ data are also used to further remove 
foreground objects.

Object detection and photometry were 
made using SExtractor version 2.1.6 (Bertin \& Arnouts 1996) 
on all six images.
The NB816-band image was chosen to detect objects. 
If more than 5 pixels whose counts were above the 2 $\sigma_{\rm sky}$ 
were connected, they were regarded as an object.
In total, 82,212 objects were detected down to the $3\sigma$ 
limiting magnitude of NB816.
For each object detected in the NB816 image, 
a $2''$-diameter magnitude was measured for 
each passband to derive the colors of each object.  
We adopted MAG$\_$AUTO for the total NB816 magnitude.
All the observed magnitudes and colors are corrected 
for Galactic absorption using the dust maps 
of Schlegel, Finkbeiner, \& Davis (1998).

%
%

\subsection{Spectra with FOCAS and DEIMOS}

We made spectroscopic observations of 63 objects selected from 
the NB816-detected catalog with FOCAS (Kashikawa et al. 2002) 
on the Subaru Telescope and DEIMOS (Faber et al. 2003) 
on the Keck II Telescope.
All 63 objects are NB816-excess objects with 
$i'-NB816 \ge 1.0$ and $NB816 \le 26.0$.
We first assigned slitlets to objects either with $R-z' \ge 1.0$ 
or undetected in $R$; they are likely LAE candidates.
We then included objects irrespective of $R-z'$ color 
in the target list if slitlets were available, 
in order to evaluate the completeness and contamination 
of the color-selected sample of LAE candidates 
to be constructed in Section 4.

%
%

\subsubsection{FOCAS Spectroscopy}

A total of 39 objects were observed with FOCAS in 2002 and 2004 
in the multi-slit mode.
Among them, 
three objects were observed on UT 2002 June 7 and 9 
with a 300 lines mm$^{-1}$ grating and an O58 order-cut filter 
using two masks.
The typical spectral coverage was 5400 \AA\ -- 10000 \AA\ 
with a pixel scale of 0.10 \AA\ pixel$^{-1}$. 
The use of $0.''8$ slits gave a resolution of $9.5$ \AA\ at 8150 \AA.
The spatial resolution was $0.''3$ pixel$^{-1}$ by 3-pixel 
on-chip binning. 
The integration time per mask was 12600 seconds.
The sky condition was good with a seeing size 
of $0.''4$ -- $1.''0$.
We also obtained spectra of the standard stars Hz 44 and 
Feige 34 for flux calibration.
The data were reduced in a standard manner.

The remaining 36 objects were observed on UT 2004 April 24-27 
with the same type of grating and order-cut filter.
Six masks were used in total.
The use of $0.''6$ slits gave a resolution of $7.1$ \AA\ at 8150 \AA.
The spatial resolution was $0.''3$ pixel$^{-1}$ by 3-pixel 
on-chip binning. 
The integration time per mask was 12000-16800 seconds.
The sky condition was good with a seeing size 
of $0.''4$ -- $0.''8$.
The standard stars Hz 44 and Feige 34 were observed 
for flux calibration.
The data were reduced in a standard manner.

%
%

\subsubsection{DEIMOS Spectroscopy}

Twenty-four objects among the 63 were observed with DEIMOS 
on UT 2004 April 23 and 24.
We used four MOS masks with an 830 line mm$^{-1}$ grating 
and a GG495 order-cut filter. 
The spectral coverage was $\sim 5000$ \AA\ -- $10000$ \AA, 
with a central wavelength of $7500$ \AA\ for one MOS mask 
and $8100$ \AA\ for the other three. 
The slit width was $1.''0$, giving a spectral resolution of 
$3.97$ \AA.
The integration times of individual masks were $7000$ -- $9000$ 
seconds, and the typical seeing sizes were $0.''55$ -- $1.''0$.
Spectra of the standard stars BD$+$28d4211 and Feige 110 were 
taken for flux calibration.
The data were reduced with the spec2d pipeline.
\footnote{
This pipeline was developed at UC Berkeley with support from 
NSF grant AST-0071048.}

%
%

\section{Spectroscopic Sample of $z=5.7$ LAEs}

%
%

\subsection{Classification of Spectra by Weighted Skewness}

We classify the 63 objects with spectroscopic observations 
into LAEs and nearby emission-line objects 
using the features of emission lines.
NB816-excess objects can be either nearby 
emitters (H$\alpha$, or [O{\sc iii}], or [O{\sc ii}] emitters), 
or LAEs at $z \simeq 5.7$.
Typically, the Ly$\alpha$ emission lines of high-redshift galaxies 
are single, asymmetric lines, with the profile blueward of 1216 \AA\ 
erased almost completely by neutral hydrogen gas. 
Thus, they can be discriminated from nearby emission lines, 
which are either single, symmetric lines, or doublet lines.

Objects with doublet emission lines are easy to classify; 
they are either [O{\sc ii}] or [O{\sc iii}] emitters depending on 
the separation of the doublet lines.
Objects with single emission lines are either LAEs, or H$\alpha$ 
emitters, or [O{\sc ii}] emitters with unresolved doublet lines.
To classify single emission-line objects, 
we introduce weighted skewness, $S_W$, 
which quantifies the asymmetry of emission lines.
A similar quantity (skewness) has been applied 
by Kurk et al. (2004) to identify an LAE at $z=6.518$.

For a given one-dimensional spectrum of an emission line, 
$f(x_i)$ $(i=1,2,...,n)$, where $x_i$ is the $i$-th pixel 
along wavelength and $f(x_i)$ is the flux in the $i$-th pixel, 
the weighted skewness is defined as 
\begin{equation}
S_W = (\lambda_{10,r} - \lambda_{10,b}) \times 
      {1\over{I\sigma^3}} \sum_{i=1}^{n} (x_i - \bar{x})^3 f(x_i).
\end{equation}
\noindent
Here, $I = \sum f(x_i)$; 
$\bar{x} = {1\over{I}} \sum x_i f(x_i)$;
$\sigma = \sqrt{{1\over{I}} \sum (x_i - \bar{x})^2 f(x_i)}$;
$\lambda_{10,r}$ and $\lambda_{10,b}$ are the wavelengths 
where the flux drops to $10\%$ of its peak value at the red 
and blue sides of the emission, respectively.
In the above formula, 
${1\over{I\sigma^3}} \sum_{i=1}^{n} (x_i - \bar{x})^3 f(x_i)$ 
measures the skewness of the line, 
and $\lambda_{10,r} - \lambda_{10,b}$ measures 
the width of the line.
Since the Ly$\alpha$ emission of high-redshift galaxies tends 
to be wider than other emission lines of nearby galaxies 
in the observed frame, 
we include the factor $\lambda_{10,r} - \lambda_{10,b}$ in $S_W$ 
to enhance the difference between Ly$\alpha$ and other lines.
$S_W$ is larger for objects with higher asymmetries and/or 
larger line widths. 
Note that for an identical object, spectroscopy with 
a higher dispersion generally gives a larger $S_W$ value.
A full description of $S_W$ is found 
in Kashikawa et al. (2005). 

Ly$\alpha$ emission lines at high redshifts typically have 
large positive $S_W$ values.
On the other hand, [O{\sc iii}] and H$\alpha$ lines are nearly 
symmetric, i.e., $S_W \simeq 0$.
The $S_W$ of [O{\sc ii}] emitters is also expected to be small 
($S_W$ is measured for the doublet as a whole); 
for all resolved [O{\sc ii}] emitters in our sample, 
the $\lambda3726$ line is weaker than the $\lambda3729$ line, 
meaning a negative $S_W$ (See also Rhoads et al. 2003).
This means that $S_W$ can discriminate LAEs from [O{\sc ii}] emitters 
even when the [O{\sc ii}] doublet is not resolved in a spectrum 
due to a low spectral resolution and/or a low signal-to-noise ratio.

\begin{figure}
  \hspace{-45pt}
  \FigureFile(120mm,120mm){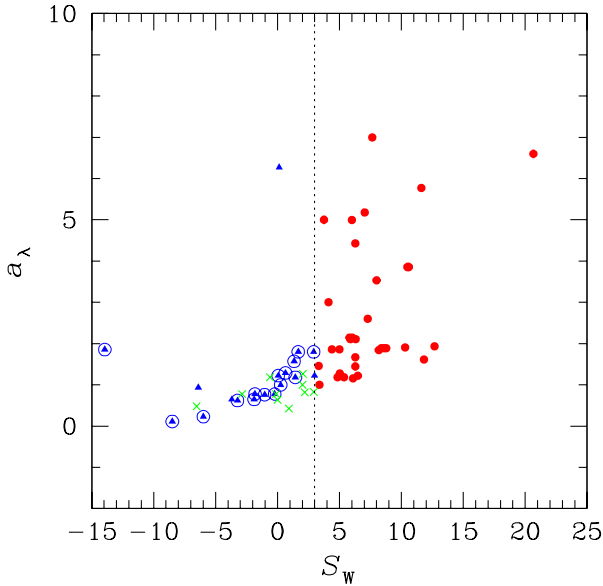}
  \vspace{-20pt}
  \caption{
         Weighted skewness of the 63 objects with spectroscopic
         observations. 
         The red circles, blue triangles, and green crosses 
         indicate LAEs, nearby objects, and unclear objects, 
         respectively.
         The objects with open blue circles have 
         multiple emission lines.
         Ordinate shows wavelength ratio, $a_\lambda$.
    \label{fig:aw_wskew}}
\end{figure}

Figure \ref{fig:aw_wskew} plots $S_W$ for the 63 objects. 
The ordinate represents, for a reference, the wavelength ratio 
($a_\lambda$) defined by Rhoads et al. (2003); 
$a_\lambda$ is defined as 
$a_\lambda = (\lambda_{10,r} - \lambda_p) /
(\lambda_p - \lambda_{10,b})$, 
where $\lambda_p$ is the wavelength of the peak flux density.
A loose correlation is found between $a_\lambda$ and 
$S_W$ in Fig. \ref{fig:aw_wskew}.
Objects marked by a large open circle are nearby emitters with 
definite classification based on their emission lines;
they are either H$\alpha$ emitters (from the presence of 
the [O{\sc iii}] line or from a very flat and strong continuum 
over the whole wavelength range), or [O{\sc iii}] emitters 
(from doublet lines at 4959 \AA\ and 5007 \AA), 
or [O{\sc ii}] emitters (from doublet lines at 3726 \AA\ and 3729 \AA).
All these nearby objects have $S_W < 3$.
A visual inspection of the 63 spectra reveals that 
most of the objects with $S_W > 3$ have 
an asymmetric emission-line shape typical of high-redshift 
Ly$\alpha$ emission.

On the basis of these results, 
we classify the spectroscopic objects into three classes as follows.
We primarily use $S_W$, but in some cases 
we also use $B$ and $V$ photometry as additional information,
since LAEs at $z=5.7$ are essentially invisible in our $B$ and $V$ data 
due to strong absorption by the IGM.
If an object has the definite feature of 
the [O{\sc ii}] or [O{\sc iii}] or 
H$\alpha$ line, it is a nearby object. 
If an object has a single emission-line with $S_W > 3$, 
it is regarded as an LAE (irrespective of the $B$ and $V$ fluxes); 
we assume that those with detected $B$ or $V$ fluxes are contaminated 
by a faint foreground object.
If, on the other hand, a single-line object with $S_W < 3$ 
is detected in either $B$ or $V$ (i.e., brighter than the $2\sigma$ 
limit in the corresponding bandpass), 
then it is classified as a nearby object.
Objects which do not satisfy either of the above three categories 
are classified as {\lq}unclear{\rq} objects; 
unclear objects are single-line emitters with 
$S_W < 3$ but invisible in both $B$ and $V$.
Among the 63 objects, 34, 19, and 10 are classified as 
LAEs, nearby objects, and unclear objects.
Among the 34 LAEs, six are detected in either $B$ or $V$.
This rather high contamination rate, $6/34=18\%$, appears 
to be accounted for by the observed surface number density 
of faint $B/V$ sources. 
We find that the fraction of the sky occupied by faint 
($m>26$) $B/V$ sources is about 10\%.
Furthermore, the probability of a faint $B/V$ source being within
$1''$ of a random position in the sky is found to be about $20\%$.
The contamination rate obtained for the spectroscopic sample 
may be compared to these rough estimates of chance projection.

\begin{figure}
  \hspace{-38pt}
  \FigureFile(105mm,105mm){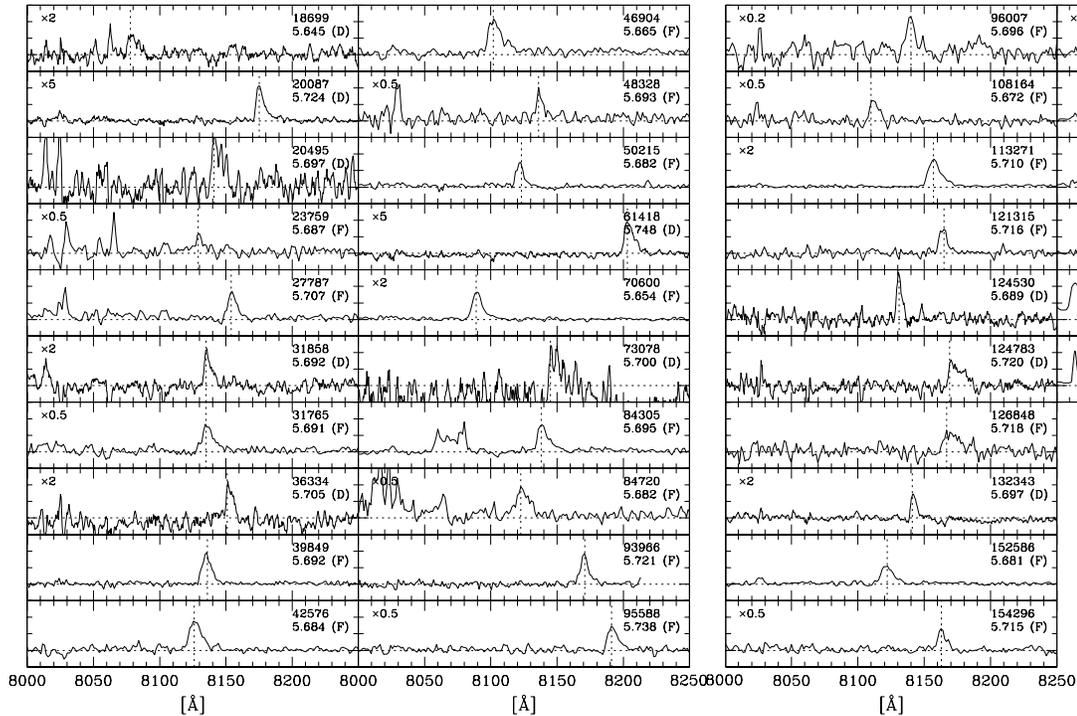}
  \vspace{-15pt}
  \caption{
         Spectra of LAEs. 
         The ID number (in the NB816-detected catalog) 
         and redshift are shown in the upper right 
         corner of each panel. {\lq}F{\rq} ({\lq}D{\rq}) in 
         parentheses indicates that the object was observed with 
         FOCAS (DEIMOS).
         The vertical dotted lines indicate the center of the 
         (single) emission line.
         The scale on the $y$ axis is marked in 
         $0.5\times 10^{-18}$ erg s$^{-1}$ cm$^{-2}$ \AA$^{-1}$; 
         for panels in which a factor is shown in the upper left 
         corner, 
         multiply the scale by this factor to obtain a correct 
         scale.
    \label{fig:spec_lae1}}
\end{figure}

\begin{figure}
  \hspace{-38pt}
  \setcounter{figure}{1}
  \FigureFile(105mm,105mm){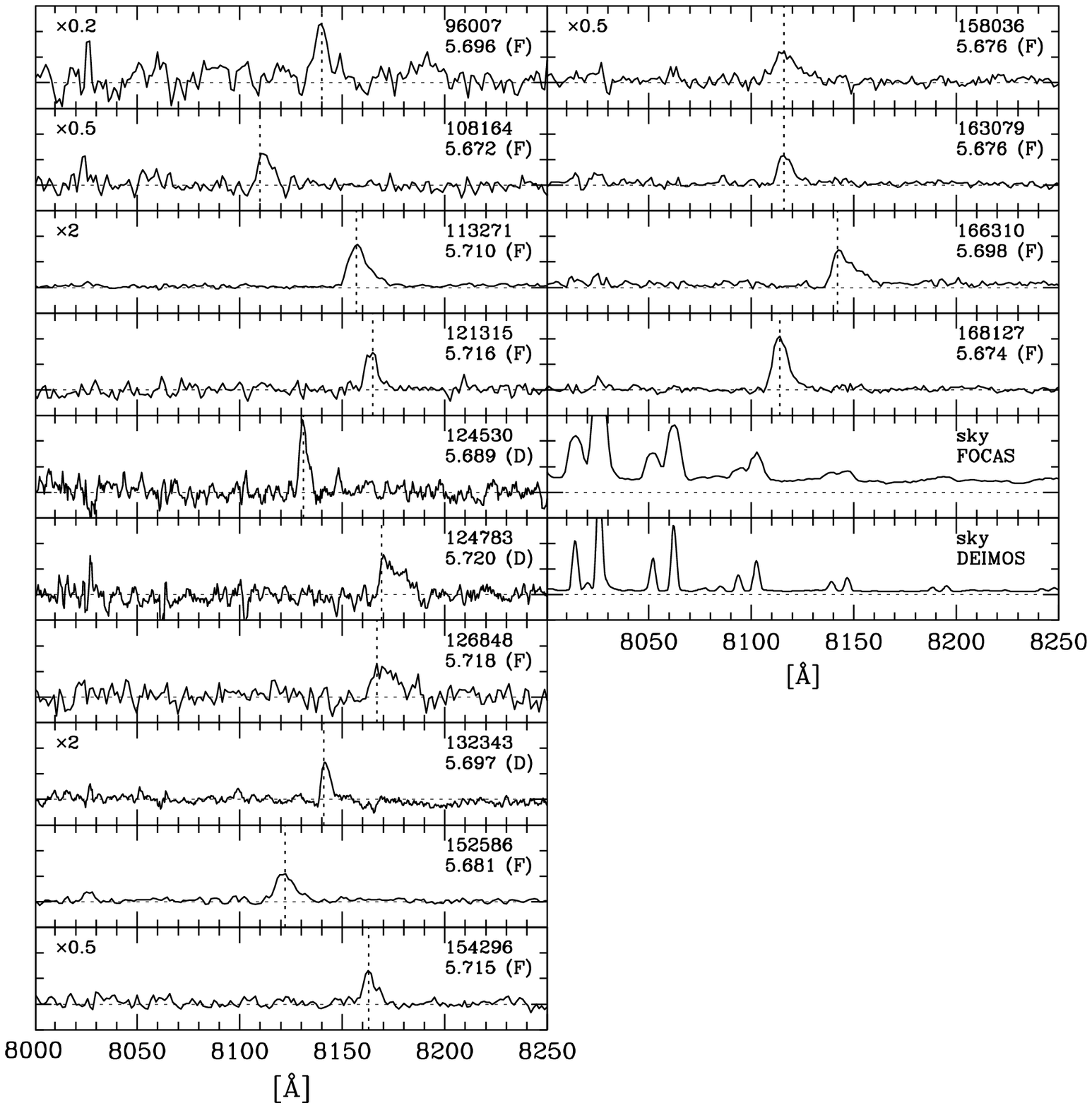}
  \vspace{-15pt}
  \caption{
         Fig. \ref{fig:spec_lae1} -- continued.
         The second bottom and bottom panels in right-hand side 
         show typical spectra of the sky background obtained 
         with FOCAS and DEIMOS, respectively. The normalization 
         is arbitrary.
    \label{fig:spec_lae2}}
\end{figure}

\begin{figure} 
  \hspace{-38pt}
  \FigureFile(105mm,105mm){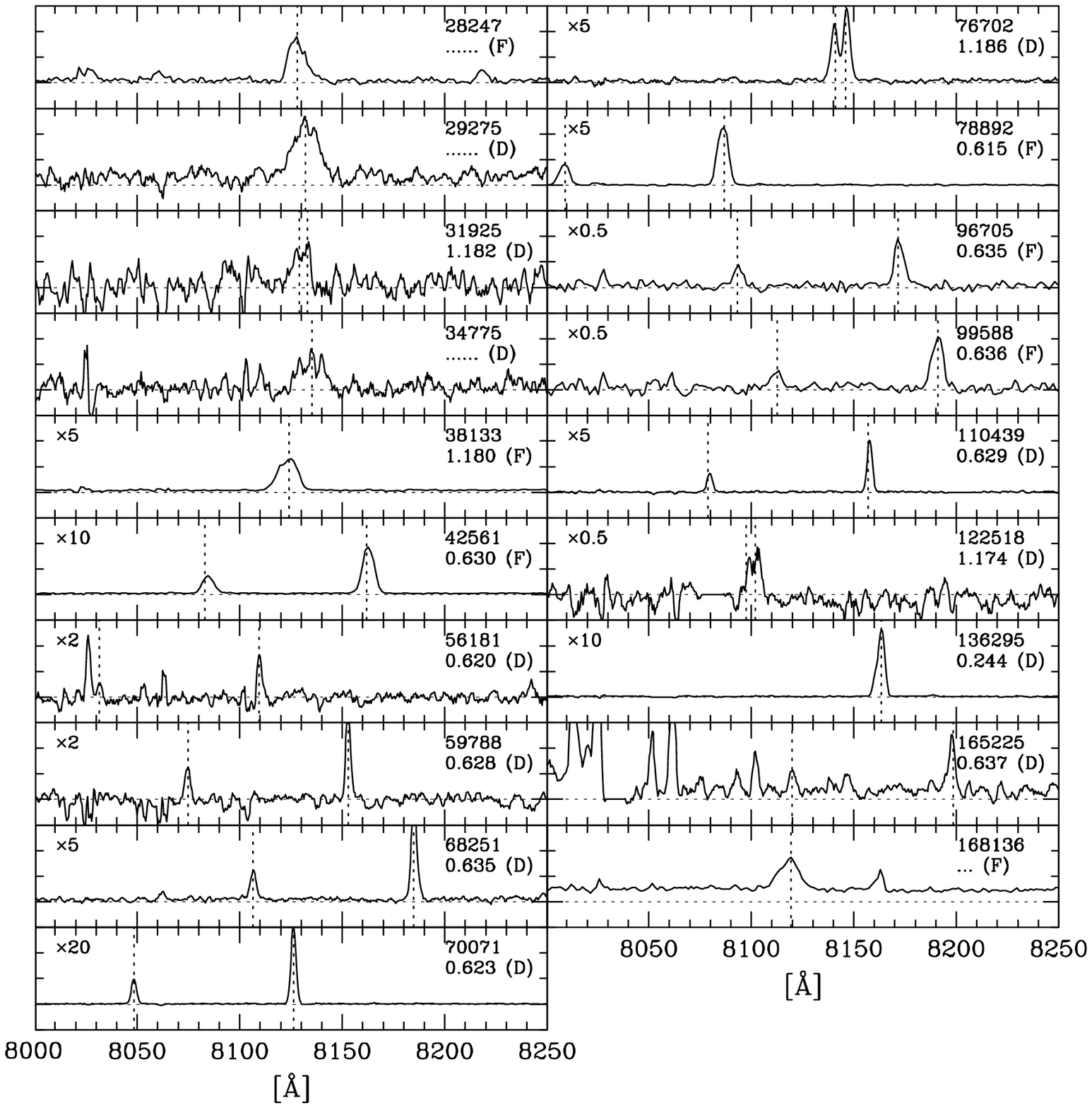}
  \vspace{-15pt}
  \caption{
         Spectra of nearby objects. 
         The ID number (in the NB816-detected catalog)
         and redshift are shown in the upper right 
         corner of each panel. {\lq}F{\rq} ({\lq}D{\rq}) in 
         parentheses indicates that the object was observed with 
         FOCAS (DEIMOS).
         The vertical dotted lines indicate the identified 
         emission lines (H$\alpha$ or 
         [O{\sc iii}] or [O{\sc ii}]).
         Identification are unsuccessful 
         for \#28247, \#29275, \#34775, and \#168136; 
         these are classified as nearby objects because of 
         a small skewness ($S_W<3$) and a bright $B$ or $V$ magnitude.
         The scale on the $y$ axis is marked in 
         $1\times 10^{-18}$ erg s$^{-1}$ cm$^{-2}$ \AA$^{-1}$; 
         for panels in which a factor is shown in the upper left 
         corner, 
         multiply the scale by this factor to obtain a correct 
         scale.
    \label{fig:spec_nearby}}
\end{figure}

\begin{figure}
  \hspace{40pt}
  \FigureFile(105mm,105mm){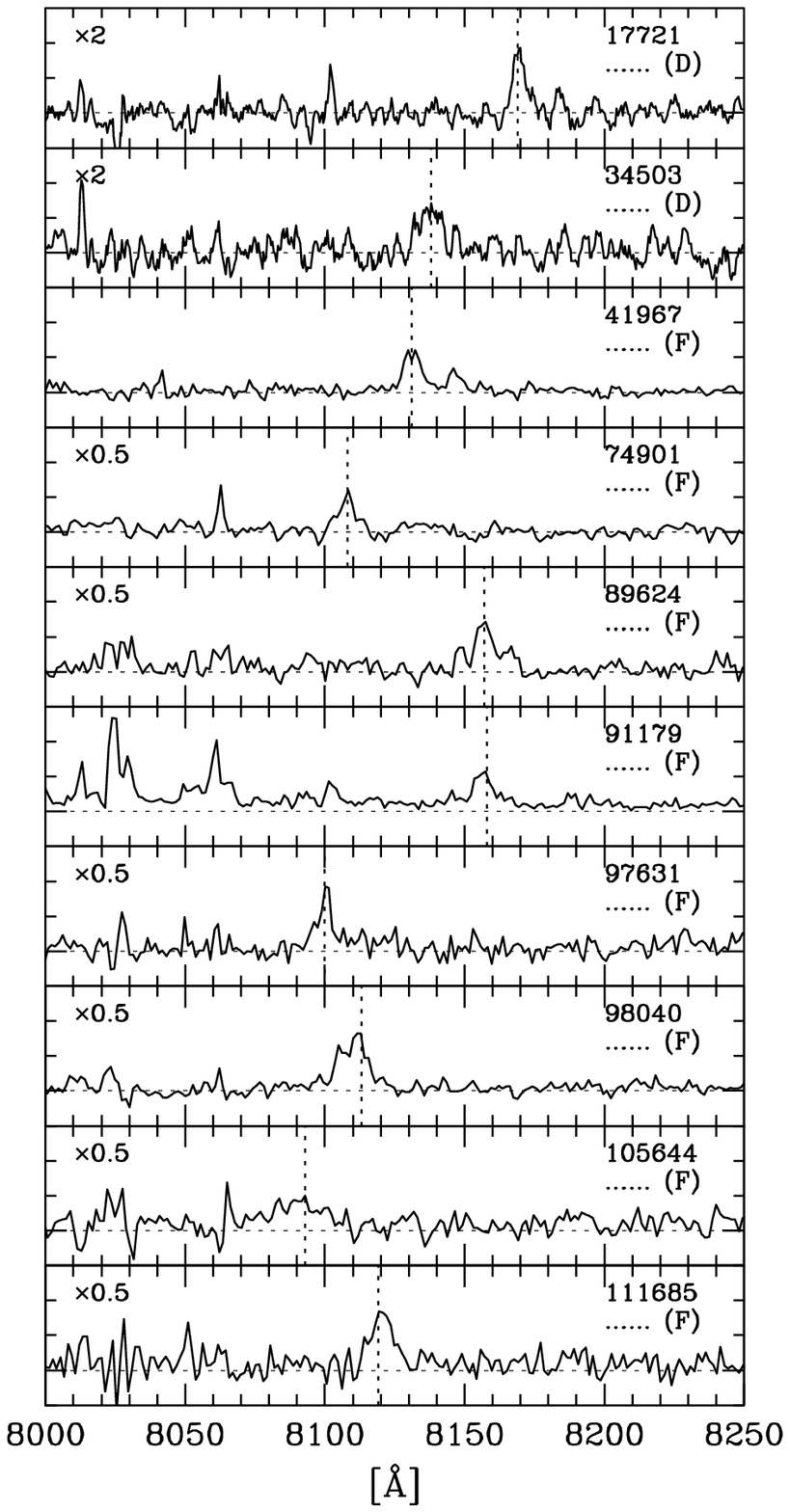}
  \vspace{-15pt}
  \caption{
         Spectra of unclear objects. 
         The ID number (in the NB816-detected catalog)
         is shown in the upper right 
         corner of each panel. {\lq}F{\rq} ({\lq}D{\rq}) in 
         parentheses indicates that the object was observed with 
         FOCAS (DEIMOS).
         The vertical dotted lines indicate the position of 
         the single emission line. 
         The scale on the $y$ axis is marked in 
         $0.5\times 10^{-18}$ erg s$^{-1}$ cm$^{-2}$ \AA$^{-1}$; 
         for panels in which a factor is shown in the upper left 
         corner, 
         multiply the scale by this factor to obtain a correct 
         scale.
    \label{fig:spec_unclear}}
\end{figure}

Table \ref{tab:obj_spec} summarizes the photometry and 
spectroscopy of the 63 objects.
Figures \ref{fig:spec_lae1} -- \ref{fig:spec_unclear} 
show the spectra of the LAEs, nearby objects, and unclear objects, 
respectively.
The luminosity functions and star formation properties of 
the line-emitting galaxies at $z \ltsim 1.2$ identified 
in our spectroscopy will be discussed 
in Ly et al. (2006, in preparation).

We infer that some portion of the unclear objects 
may be LAEs in reality. 
This is because (i) not all LAEs may have as large asymmetry as 
the confirmed LAEs
and (ii) since the resolution of the FOCAS spectra 
is significantly poorer than that of the DEIMOS spectra, 
the absolute value of $S_W$ of an object based on a FOCAS spectrum 
will be smaller than that based on a DEIMOS spectrum.
In this sense, the criterion of $S_W > 3$ 
is conservative in terms of the selection of LAEs.
Among the ten unclear objects, eight are FOCAS objects.
When estimating the selection completeness of the photometric sample 
of $z=5.7$ LAEs in Section 4, we take account of 
the extreme possibility that all of the unclear objects 
are actually LAEs.

\begin{figure}
  \hspace{-45pt}
  \FigureFile(140mm,140mm){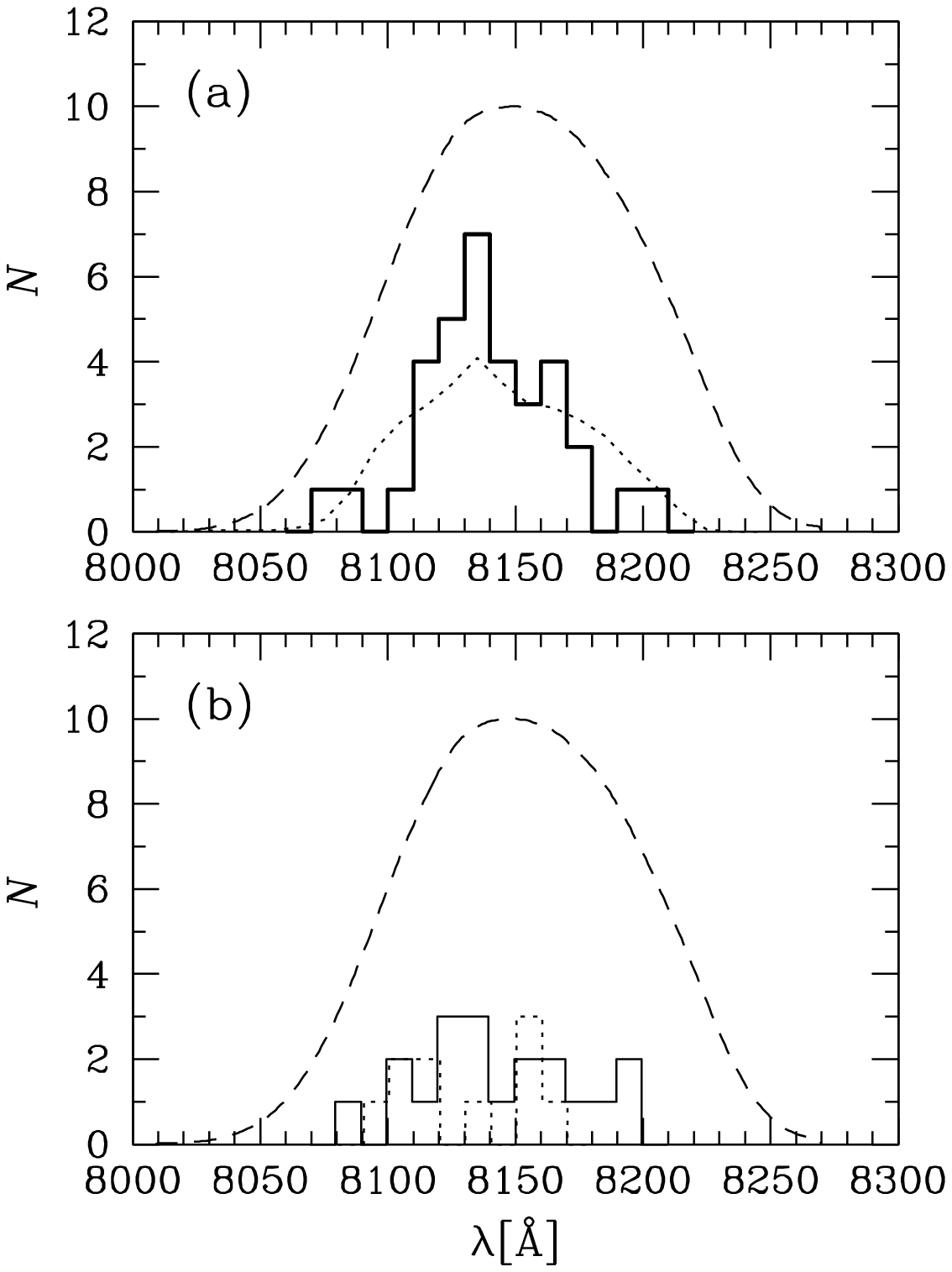}
  \vspace{-70pt}
  \caption{
         Distributions of emission line wavelengths for 
         objects with spectroscopic observations.
         {\it Panel (a)}. The distribution for the 34 LAEs is 
         shown by the thick solid histogram.
         {\it Panel (b)}. The distributions for the 19 nearby 
         objects (thin solid histogram) and the 10 unclear 
         objects (dotted histogram).
         The dotted curve in panel (a) represents 
         the distribution of model LAEs with $NB816 \le 26.0$ 
         in a mock sample (See text).
         In both panels, the dashed curve corresponds to the 
         response function of the NB816 band.
    \label{fig:hist_lam}}
\end{figure}

Figure \ref{fig:hist_lam} shows the distributions of emission-line 
wavelengths (the central value if the emission is a doublet) 
for all of our objects.
The thick solid histogram in panel (a) represents 
the distribution for the 34 LAEs, 
and panel (b) plots the distributions for nearby objects 
(thin solid histogram) and unclear objects (dotted histogram).
In both panels, the dashed curve corresponds to the response 
function of the NB816 band.
It is found that while the distributions of nearby and 
unclear objects are statistically consistent with 
being symmetric around the central wavelength of the bandpass, 
the LAEs' distribution is skewed toward shorter wavelengths.
A natural explanation for this skewed distribution is as follows.
LAEs with higher redshifts are on average fainter 
in NB816 (even when the Ly$\alpha$ luminosity is fixed) 
because of a lower fraction of the continuum emission redward of 
$\lambda_{\rm rest} = 1216$ \AA\ falling into the NB816 bandpass.
Hence they tend to be excluded from a list of spectroscopic targets. 

The dotted curve in panel (a) represents 
the distribution of modeled LAEs with $NB816 \le 26.0$ 
in a mock sample; 
this sample is constructed 
based on the same selection criteria as for the observed sample 
(see the next section) from a large catalog
of LAEs generated by Monte Carlo simulations 
for a range of LAE equivalent widths and 
luminosity function parameters (see Section 5). 
In the simulations, 
LAEs are distributed uniformly in comoving space 
over the redshift range $5.6 < z < 5.8$, 
which corresponds to an observed Ly$\alpha$ wavelength 
of 8020 to 8270 \AA.
The thick histogram is found to agree fairly well 
with the dotted curve, suggesting that prominent  
large-scale structures along the line of sight 
are not present in our LAE sample.

Hu et al. (2004) also found a skewed distribution in their 
$z=5.7$ LAEs selected from NB816 data in the SSA22 field, 
and argued that the true center in wavelength of the NB816 filter 
may be shifted toward shorter wavelengths by about 25 \AA\ 
than the measured one. 
Although we cannot completely rule out this possibility, 
the wavelength distribution of our sample is explained 
without assuming such a shift.

%
%

\subsection{Confirmed Sample of $z=5.7$ LAEs}

We have 34 LAEs with spectroscopic confirmation.
Their distribution in the $i'-NB816$ vs $NB816$ plane is 
slightly biased toward brighter NB816 magnitudes and 
redder $i'- NB816$ colors 
with respect to the photometric sample, 
except for three objects of $i'- NB816 \simeq 1.2$.
Among the 34 objects, the faintest in NB816 is $25.89$ mag.

\begin{figure}
  \hspace{-45pt}
  \FigureFile(120mm,120mm){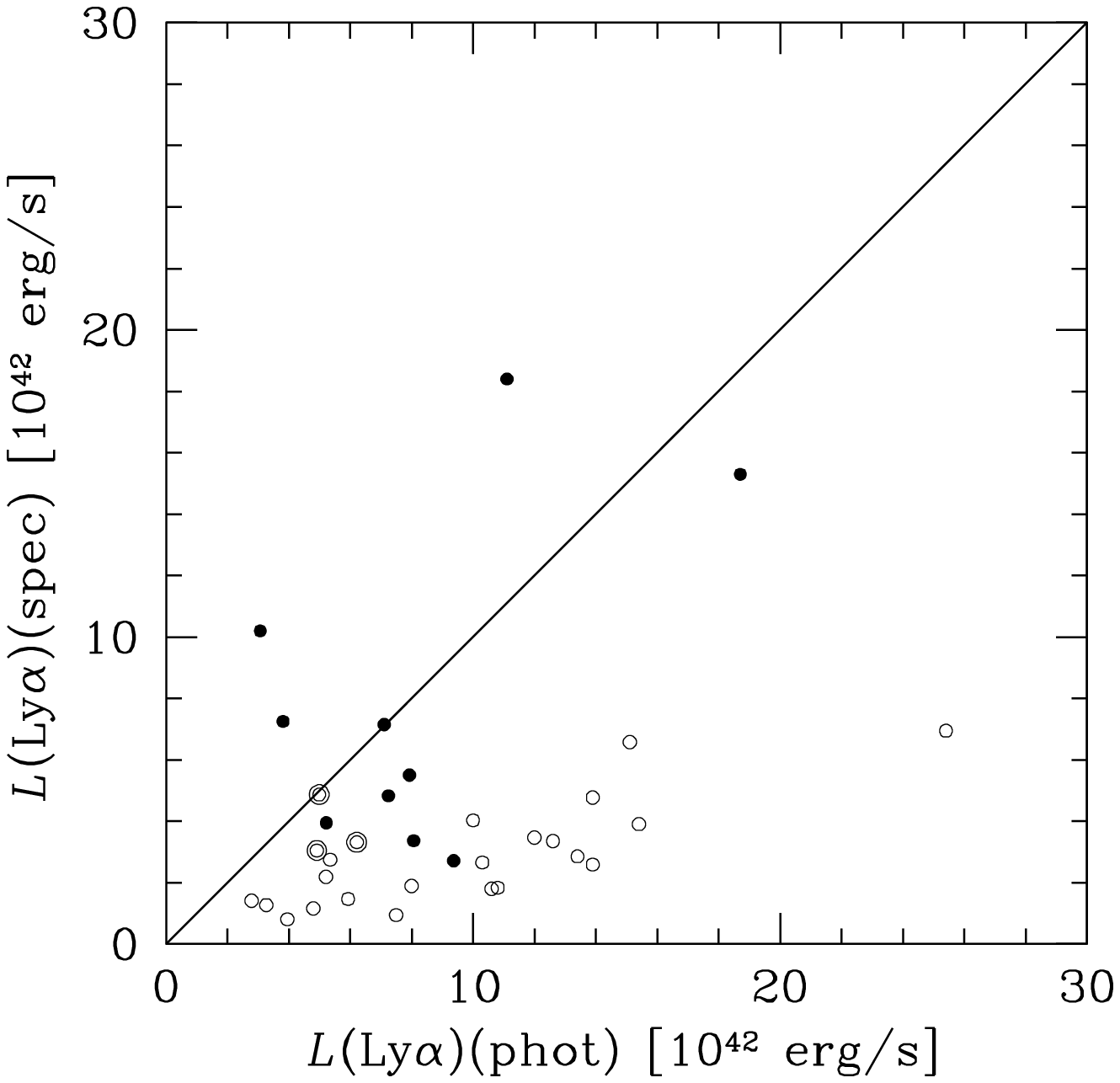}
  \vspace{-30pt}
  \caption{
         Ly$\alpha$ luminosities measured from the spectra 
         are plotted against those based on the photometric data 
         for the 34 LAEs with spectroscopic confirmation.
         The filled and open circles indicate objects observed 
         with DEIMOS and FOCAS, respectively. 
         Three FOCAS objects observed through $0.''8$ slitlets 
         are marked with large open circles. 
    \label{fig:comp_LLyA}}
\end{figure}

We measure the flux and FWHM of the Ly$\alpha$ line 
from the spectrum of each object.
In this paper, however, we adopt for Ly$\alpha$ fluxes 
the values calculated from the imaging data 
by subtracting the continuum flux measured from the $z'$-band image 
from the flux falling into the NB816 band, assuming 
$f_\nu$ = constant; spectroscopic 
redshifts are used to determine the precise position of 
the Ly$\alpha$ line in the NB816 response function.
The reason for not using the Ly$\alpha$ fluxes directly
is that our multi-slit 
spectroscopy probably lost a non-negligible amount of 
Ly$\alpha$ photons, 
because of the relatively narrow slitlets 
($0.''6$ and $0.''8$ for FOCAS and $1.''0$ for DEIMOS) 
compared with the typical PSF sizes, 
and because the Ly$\alpha$ emission could be extended spatially.
Figure \ref{fig:comp_LLyA} shows Ly$\alpha$ luminosities 
measured from the spectra against those from photometry 
(See Subsection 5.2 for the calculation of Ly$\alpha$ luminosities 
based on the photometric data).
While little systematic difference is seen between the luminosities 
from the DEIMOS spectra and from the photometry, 
the FOCAS spectra give luminosities only about one third those 
measured from the photometric data. 
Among the FOCAS objects, three were observed with a $0.8''$ slitlet, 
and their 
$L({\rm Ly}\alpha({\rm spec}))/L({\rm Ly}\alpha({\rm phot}))$ ratio 
is $70\%$ on average.
These results are qualitatively consistent with narrower slitlets 
losing a larger fraction of Ly$\alpha$ photons. 
It is difficult to correct $L({\rm Ly}\alpha({\rm spec}))$ for this 
effect, since the seeing size varied during the exposures and 
since different objects have different intrinsic sizes.
For the same reason, we also measure the equivalent widths of 
the 34 objects from the photometric data, as described 
in Subsection 5.2.

%
%

\section{Photometric Sample of $z=5.7$ LAEs}

We construct a photometrically selected LAE 
sample from the imaging data. 
Objects located in low-quality regions of the images 
are excluded from the selection process to obtain an LAE sample 
of uniform detection.
The low-quality regions include areas around very bright stars, 
the four edges of the images, and the south-east region 
which corresponds to the field of view of the CCD whose quantum 
efficiency is only about two thirds those of the other nine CCDs.
The effective area is thus reduced to 725 arcmin$^2$.

\begin{figure}
  \hspace{0pt}
  \FigureFile(90mm,90mm){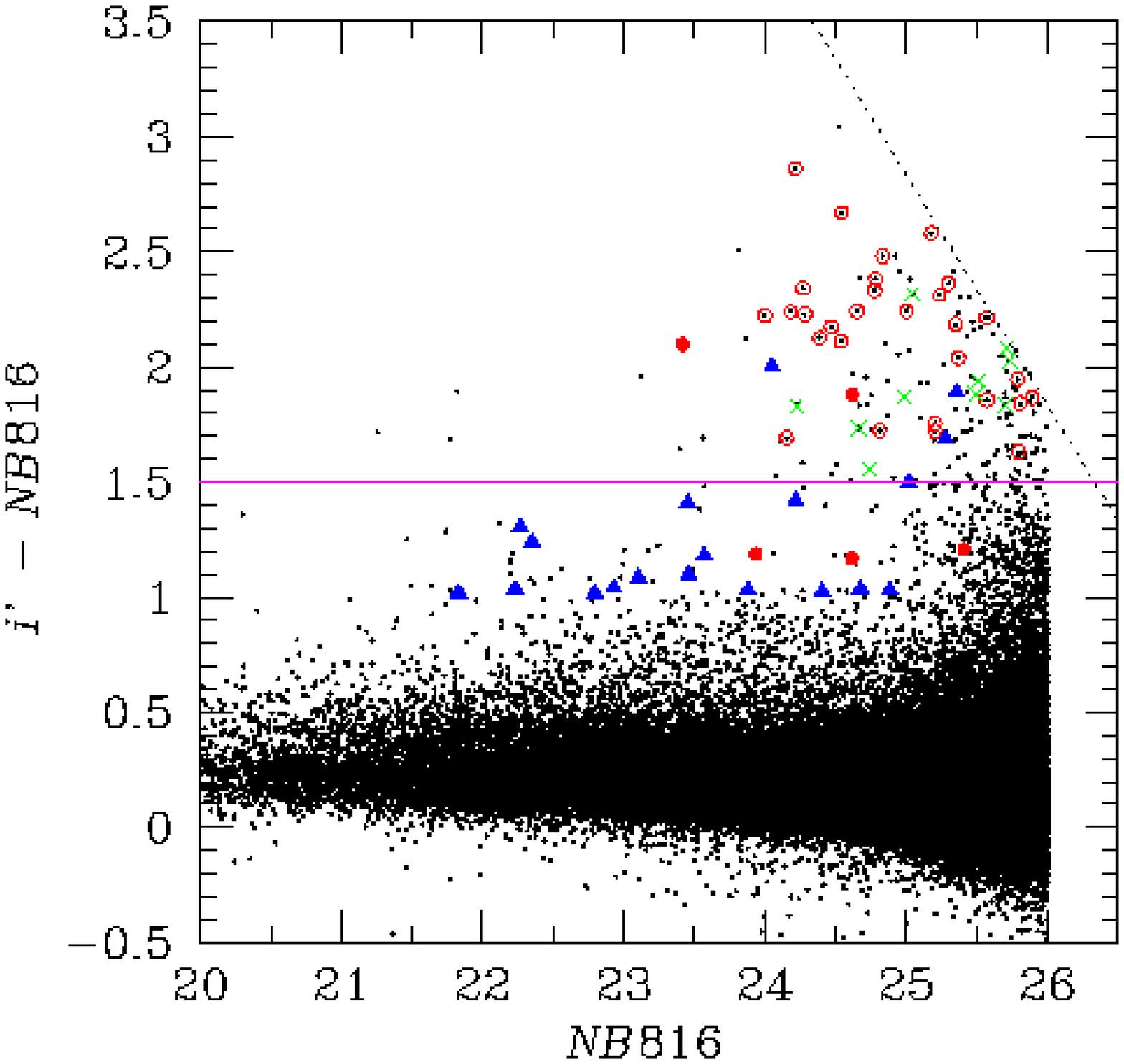}
  \vspace{-10pt}
  \caption{
         $i'- NB816$ plotted against $NB816$ for objects 
         with $NB816 \le 26.0$.
         The large symbols indicate objects with spectroscopic 
         observations: 
         red circles, blue triangles, and green crosses 
         indicate LAEs 
         ($N=34$), nearby objects ($19$), and unclear objects ($10$), 
         respectively.
         For LAEs, open circles mean $R$ magnitudes fainter than 
         the $2\sigma$ magnitude.
         The magenta line shows the threshold of $i'- NB816$ 
         adopted for the selection of LAEs: $i'- NB816 \ge 1.5$ 
         (see text).
         The dotted line implies the $i'$-band $2\sigma$ limit.
    \label{fig:colormag}}
\end{figure}

Figure \ref{fig:colormag} plots $i'- NB816$ color 
against NB816 magnitude for all objects with $NB816 \le 26.0$ 
($N=61,919$). 
The vast majority of the objects are distributed around 
$i'- NB816 \sim 0.2$; their $i'- NB816$ distribution 
spreads as NB816 magnitude goes fainter, 
mainly due to the increase in photometric errors.
Objects with large $i'- NB816$ colors are emission-line 
objects (or some kind of late-type stars with $i'- NB816 > 1$), 
part of which are LAEs at $z=5.7$.
The large symbols indicate objects with spectroscopic observations: 
red circles, blue triangles, and green crosses 
indicate LAEs ($N=34$), 
nearby objects ($19$), and unclear objects ($10$), respectively.
For LAEs, open circles mean 
$R$ magnitudes fainter than the $2\sigma$ magnitude ($N=29$).
The magenta line shows the threshold of $i'- NB816$ 
adopted for the selection of LAEs: $i'- NB816 \ge 1.5$ (see below).
The dotted line implies the $i'$-band $2\sigma$ magnitude.
The $i'$-band magnitudes of objects 
undetected in $i'$ (i.e., fainter than the $2\sigma$ magnitude) 
have been replaced with the $2\sigma$ magnitude.

\begin{figure}[h]
  \hspace{0pt}
  \FigureFile(90mm,90mm){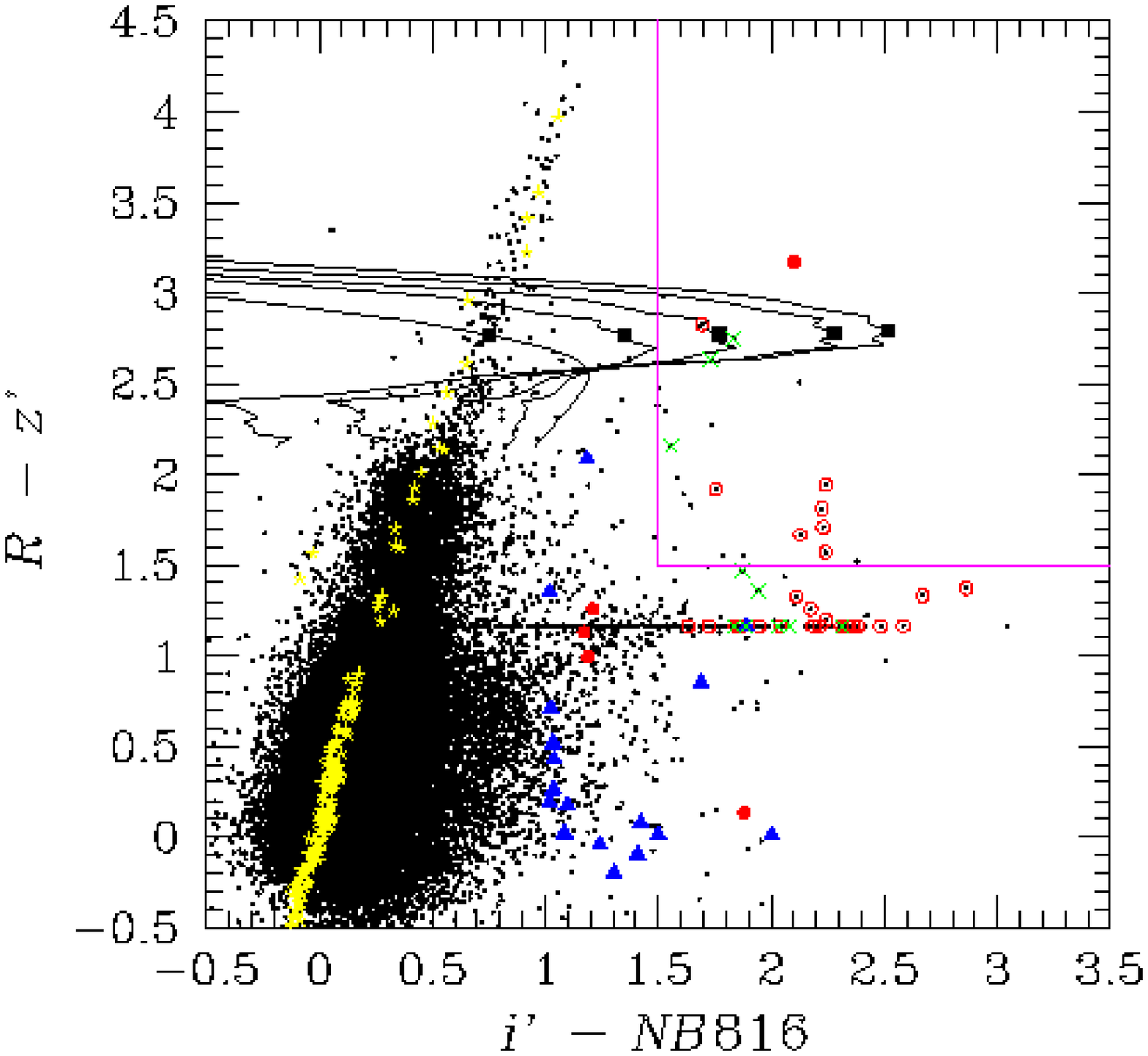}
  \vspace{-10pt}
  \caption{
         Distribution of objects 
         in the $R-z'$ vs $i'- NB816$ plane.
         The dots indicate  objects with $NB816 \le 26.0$.
         If an objects is undetected (i.e., fainter than the $2\sigma$ 
         limit) in a given bandpass, its magnitude in that bandpass 
         is replaced with the $2\sigma$ magnitude.
         Objects on the $R-z'=1.16$ line are those undetected in 
         both $R$ and $z'$.
         The red circles represent the spectroscopically confirmed 
         LAEs ($N=34$); open circles show that they are undetected 
         in $R$.
         The blue triangles and green crosses 
         indicate nearby objects ($N=19$) 
         and unclear objects ($N=10$), respectively.
         The yellow stars correspond to the 175 Galactic stars 
         given in Gunn \& Stryker's (1983) spectrophotometric atlas.
         The black solid curves show the tracks of model LAEs 
         over a redshift range of $5.5 \le z \le 5.9$ 
         with four different intrinsic EW values of Ly$\alpha$ 
         in the rest frame: 
         $W{\rm_{0}^{i}} = 0$, $20$, $50$, $150$, 
         and $300$ \AA\ from left to right;
         filled squares correspond to $z=5.7$.
         The magenta solid lines indicate the selection boundary 
         (in the case of $m_{\rm R}<m_{\rm R}(2\sigma$)).
    \label{fig:twocolor}}
\end{figure}

Figure \ref{fig:twocolor} shows the distribution of 
objects in the $R-z'$ vs $i'- NB816$ plane.
In this figure, if an object is undetected 
(i.e., fainter than the $2\sigma$ magnitude) 
in a given bandpass, its magnitude in that bandpass is replaced 
with the $2\sigma$ magnitude.
Objects on the $R-z'=1.16$ line are those undetected in 
both $R$ and $z'$.
More than half ($18/34$) of the LAEs are invisible in $z'$.
The red circles represent the spectroscopically confirmed LAEs 
($N=34$); 
open circles show that they are undetected in $R$ ($N=29$).
The blue triangles and green crosses indicate nearby objects ($19$) 
and unclear objects ($10$), respectively.
The yellow stars correspond to the 175 Galactic stars 
given in Gunn \& Stryker's (1983) spectrophotometric atlas.
The spectroscopically confirmed LAEs are found to be separate 
fairly well from the nearby objects.
On the other hand, the unclear objects are located 
in a similar area to the LAEs; 
this may suggest that a significant fraction of the unclear objects 
are LAEs in reality.

The black solid curves show the tracks of model LAEs 
over a redshift range of $5.5 \le z \le 5.9$ 
with five different intrinsic EW values of Ly$\alpha$ 
in the rest frame: $W{\rm_{0}^{i}} = 0$, $20$, $50$, $150$, 
and $300$ \AA\ from left to right;
filled squares correspond to $z=5.7$.
In this paper, we assume that the intrinsic profile of 
Ly$\alpha$ emission is symmetric centered on zero velocity, 
and that foreground H{\sc i} gas reduces intrinsic 
EWs $50\%$ (blue half).
Thus, we set the rest-frame apparent 
equivalent width after truncation by H{\sc i} gas 
to be $W{\rm_{0}^{a}} = 0.5 \times W{\rm_{0}^{i}}$; 
the quantity measured from the data is $W{\rm_{0}^{a}}$.
Note that the assumption of $50\%$ reduction 
has not been verified for LAEs at $z \sim 5.7$ 
and may be an oversimplification.
As stated in Dawson et al. (2004), 
the dust content and detailed kinematics of the galaxy ISM 
appear to play a significant role 
in determining the emergent Ly$\alpha$ profile  
(e.g., Kunth et al. 1998; 
Stern \& Spinrad 1999; Mas-Hesse et al. 2003; 
Shapley et al. 2003; Ahn 2004).
LAEs at $z \sim 5.7$ may have a wide variety in the total 
absorption fraction of Ly$\alpha$ photons.

To calculate the model tracks, 
we construct the continuum spectral energy distribution of model 
galaxies using Kodama \& Arimoto (1997) stellar population synthesis 
model, adopting a Salpeter IMF, a $10^7$ yr age, and 
constant star formation.
Note that LAEs without Ly$\alpha$ emission can have 
as red $i'- NB816$ colors as $\simeq 1.2$.

Based on this figure, we select $z=5.7$ LAEs, by imposing 
the following four criteria 
on the NB816-detected multi-color catalog:
\begin{small}
\begin{eqnarray}
NB816 \le 26.0 & \hspace{5pt} {\rm (i)} 
     \nonumber \\
i'- NB816 \ge 1.5 & \hspace{5pt} {\rm (ii)} 
     \nonumber \\
\{R-z' \ge 1.5 \hspace{5pt} 
    {\rm and}\hspace{3pt}m_{\rm R}<m_{\rm R}(2\sigma)\}
     \hspace{5pt} {\rm or} \hspace{5pt} 
\{m_{\rm R} \ge m_{\rm R}(2\sigma)\} \hspace{5pt} & 
     \hspace{5pt} {\rm (iii)} 
     \nonumber \\
m_{\rm B} \ge m_{\rm B}(2\sigma) \hspace{5pt} {\rm and} 
     \hspace{5pt}
m_{\rm V} \ge m_{\rm V}(2\sigma) & \hspace{5pt} {\rm (iv)} 
     \nonumber 
\end{eqnarray}
\end{small}
Objects satisfying these four criteria simultaneously 
are regarded as LAE candidates.
The magenta solid lines in the figure indicate 
the selection boundary (in the case of $m_{\rm R}<m_{\rm R}(2\sigma$)) 
in the $R-z'$ vs $i'- NB816$ plane.
In total 89 LAE candidates are selected, 
which we call the photometric sample.

Table \ref{tab:obj_spec_num} summarizes 
a breakdown of the spectroscopic sample ($N=63$) 
in terms of the selection criteria for our photometric sample.
Thirty-nine out of the 63 objects satisfy the photometric selection 
criteria for LAEs; 28 are confirmed LAEs, one is a nearby galaxy, 
and ten are unclassified.
In the table, numbers in parentheses are the number of objects 
lying in the {\lq}high-quality{\rq} region of 725 arcmin$^2$ 
used to construct the photometric sample of LAEs.
Among the 28 confirmed LAEs, one is outside the {\lq}high-quality{\rq} 
region, and thus is not included in the photometric sample.
Thus, the 27 spectroscopically confirmed LAEs plus 
62 photometrically selected candidates comprise 
our photometric sample of 89 LAEs.
When discussing the Ly$\alpha$ equivalent-width distribution 
of the confirmed LAEs in subsection 5.2, we use the 28 LAEs 
which pass the photometric selection criteria. 
On the other hand, the contamination and completeness 
of the photometric sample are estimated using 
the 56 objects lying in the high-quality region.

The criterion of $i'- NB816 \ge 1.5$ picks up LAEs 
with rest-frame equivalent widths of 
$W{\rm_{0}^{i}} \gtsim 20$ \AA (or $W{\rm_{0}^{a}} \gtsim 10$ \AA). 
We adopt this criterion in order 
(i) not to select objects with very weak or no Ly$\alpha$ 
emission, 
and (ii) to avoid a high contamination by foreground objects whose 
$i'- NB816$ colors happen to exceed the criterion 
due to photometric errors.
This lower limit to EWs in our selection  
is close to those set by Ajiki et al.  (2003), 
$W{\rm_{obs}} \ge 180$ \AA\ ($W_0 \ge 27$ \AA), 
and by Rhoads \& Malhotra (2001), 
$W{\rm_{obs}} \ge 75$ \AA\ ($W_0 \ge 11$ \AA). 
However, our lower limit to EWs 
is slightly higher than that used by Hu et al. 
(2004), $I- NB816 \ge 0.7$ -- $1.0$ depending on the 
$R-z'$ color of the object. 
For LAEs at $z=5.7$, Hu et al.'s criterion corresponds to 
$i'- NB816 \gtsim 1.1$ -- $1.4$.
However, we find in our data that changing the criterion 
between $i'- NB816 \ge 1.0$ and $ \ge1.5$ does not significantly 
change the total number of candidates selected, if the selection 
is limited to objects brighter than $NB816=25.5$; 
for objects fainter than this, the contamination 
by foreground objects due to photometric errors is expected to be high 
at $i'- NB816 \le 1.5$ (see Fig. \ref{fig:colormag}), 
and thus we cannot make a reliable comparison of the number of LAEs.
For $NB816 < 25.5$, the condition $i'- NB816 \ge 1.0$ 
leaves 66 objects while $i'- NB816 \ge 1.5$ selects 53. 
Considering that the $i'- NB816 \ge 1.0$ candidates will 
include a non-negligible fraction of foreground objects 
even at $NB816 < 25$, 
the difference in the number of true LAEs between 
the two $i'- NB816$ limits will be small.

We estimate the contamination and completeness of 
this photometric sample on the basis of 
the spectroscopic observations.
The completeness of object detection, which is another completeness 
to be considered in calculating the LF, 
is discussed in the next section.
As shown in Tab. \ref{tab:obj_spec_num}, 
56 out of the 63 objects with spectroscopic observation
are from the region used to construct the photometric sample
(725 arcmin$^2$).
Among them, 36 satisfy the above criteria for LAE selection;
27 are LAEs, one nearby, and eight unclear objects.
If all of the unclear objects are LAEs in reality,
the contamination rate is calculated to be $1/36$.
On the contrary, if all of the unclear objects are nearby objects,
then the contamination is $(1+8)/36$.
The average of these two extremes, $14\%$, is regarded as an estimate
of contamination.
The completeness is estimated as follows.
Among the 56 spectroscopic objects lying in the region used to
construct the photometric sample, 32 are confirmed LAEs and
eight are unclear.
Then, 27 out of the 32 LAEs and all eight unclear objects
satisfy the photometric selection criteria for LAEs.
Thus, the completeness is calculated as $(27+8)/(32+8)$
if all of the unclear objects are actually LAEs.
It is reduced to $27/32$ if all of the unclear objects
are nearby objects.
The simple average of these two values is $86 \%$.

These contamination and completeness estimates 
will not be so accurate, since the spectroscopic targets 
have not been chosen very uniformly in the two-color plane.
When deriving the LFs of LAEs in the next section,
we do not correct them for either completeness or contamination,
because the completeness and contamination corrections using
the values obtained above almost cancel;
the combined factor is $(1-0.14)/0.86 = 1.00$.

%
%

\section{Results and Discussion}

In this section, we derive the Ly$\alpha$ LF and 
the far UV LF of $z=5.7$ LAEs from the photometric sample, 
and examine the distribution of equivalent widths 
using the spectroscopic sample.
The sky distribution of LAEs in the photometric and spectroscopic 
samples is also discussed.

\subsection{Ly$\alpha$ Luminosity Function}

\subsubsection{Calculation of the Luminosity Function}

Previous studies derived the Ly$\alpha$ LF 
by simply dividing the observed number counts of 
LAE candidates in a given narrow band by the effective survey 
volume defined as the FWHM of the bandpass times the area 
of the survey.
Although this procedure is accurate when the narrow-band filter 
has an ideal, boxcar shape, the shapes of actual filters used 
in LAE surveys are rather close to a triangle.
In such cases, the narrow-band magnitude of LAEs of 
a fixed Ly$\alpha$ luminosity varies largely as a function 
of redshift.
The selection function of LAEs in terms of equivalent width 
also changes with redshift; the minimum EW value corresponding 
to a given (fixed) narrow-band excess, such as $i'- NB816$, 
becomes larger when the redshift of the object goes away 
from the central redshift of the bandpass.

In order to avoid such uncertainties, we perform the following 
Monte Carlo simulations to find the best-fit Schechter parameters 
for the Ly$\alpha$ LF of $z=5.7$ LAEs.
First, we generate LAEs according to a given set of the Schechter 
parameters ($\alpha, \phi^\star, L^\star$) uniformly 
in comoving space over the redshift range $5.6 \le z \le 5.8$.
Then, we {\lq}observe{\rq} these LAEs in the NB816, $R$, $i'$, 
and $z'$ bands, add to their flux densities photon noise corresponding 
to the actual observation, and select LAEs 
by the same criteria as for selecting the actual LAEs.
Thus, we have a mock catalog of LAEs for the given set of 
($\alpha, \phi^\star, L^\star$).
Finally, we compare the NB816-band number counts derived from the 
mock catalog with the observed counts, and compute the $\chi^2$ value.
By performing this set of simulations over a wide range 
of the Schechter parameters, we find the best-fit parameters.

Three points should be noted in these simulations.
First, in the actual simulations we search for the best-fit 
$\phi^\star$ and $L^\star$ values for only three values of 
$\alpha$, $-1, -1.5, -2$, 
since the magnitude range of the observed number counts 
is not wide enough to place strong constraints 
on all three parameters simultaneously.
These three $\alpha$ values are chosen 
following Malhotra \& Rhoads (2004).
The difference in $\chi^2$ among the three $\alpha$ values 
is found to be insignificant, although $\alpha=-1$ gives 
the lowest $\chi^2$ value.
We adopt the $\alpha=-1.5$ results as the fiducial set of 
the best-fit Schechter parameters.

\begin{figure}
  \hspace{-45pt}
  \FigureFile(120mm,120mm){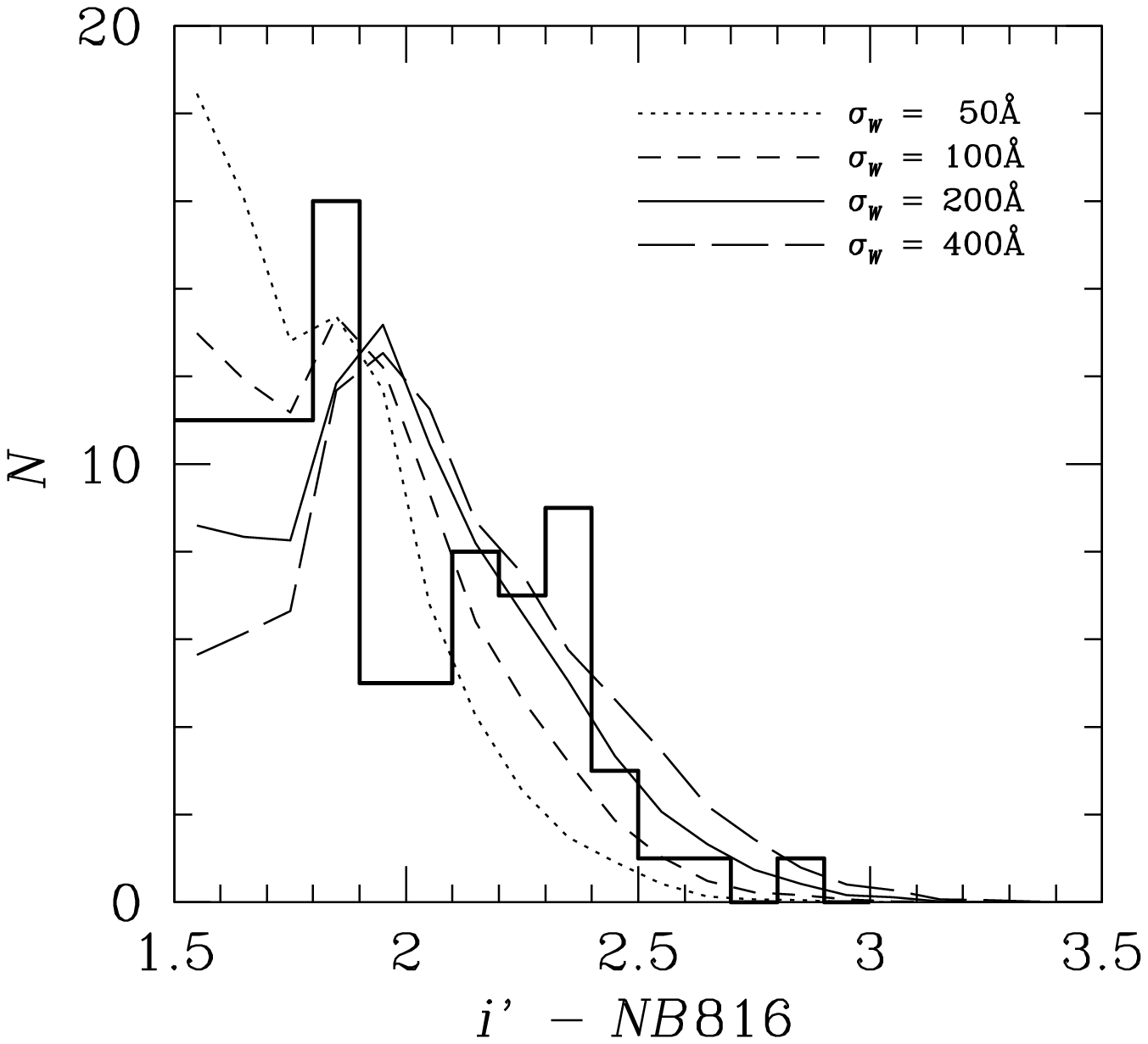}
  \vspace{-30pt}
  \caption{
         Histogram of $i'- NB816$ for the photometric sample.
         The four curves indicate the predictions from simulations 
         assuming four values for $\sigma_W$: 
         $\sigma_W = 50, 100, 200, 400$ \AA.
    \label{fig:hist_color}}
\end{figure}

Second, in order to compute flux densities in the individual 
bandpasses from a given Ly$\alpha$ luminosity, 
we need to give an equivalent width.
We assume a Gaussian distribution, 
$f(W{\rm_{0}^{i}})dW{\rm_{0}^{i}} 
\propto \exp(-W{\rm_{0}^{i}}^2/2\sigma_W^2) dW{\rm_{0}^{i}}$, 
for the probability distribution of EWs for $z=5.7$ LAEs, 
and generate objects by randomly sampling EWs from 
the $W{\rm_{0}^{i}}>0$ part of this distribution.
The value of $\sigma_W$ can be inferred 
from the observed $i'- NB816$ distribution and EW distribution.
Figure \ref{fig:hist_color} compares the predicted $i'- NB816$ 
distribution with the observation for four $\sigma_W$ values: 
$\sigma_W = 50$ \AA, 100 \AA, 200 \AA, and 400 \AA.
For each prediction, $\alpha=-1.5$ and the best-fit $L^\star$ are 
used.
It is found that $\sigma_W \approx 100$ \AA\ and $200$ \AA\ 
are preferred among the four values. 
We then calculate the equivalent width distribution 
for $\sigma_W = 100$ \AA\ and 200 \AA, 
to find that the distribution for $\sigma_W = 200$ \AA\ 
is roughly consistent with the observed distribution of the 
spectroscopic sample, while that for $\sigma_W = 100$ \AA\ 
is much skewed toward lower EWs and thus clearly 
inconsistent with the observation.
Based on these results, 
we decide to use $\sigma_W=200$ \AA\ 
to find the best-fit Schechter parameters. 
We have confirmed that using $\sigma_W=100$ \AA\ does not 
significantly change the best-fit parameters.

Third, we correct the observed number counts for detection 
completeness, since all objects are detected in the simulations 
irrespective of their magnitudes.
We estimate the detection completeness as a function of 
apparent NB816 magnitude, by distributing pseudo objects 
on the NB816 image after adding photon noise, 
detecting them by SExtractor, and computing the detection fraction.
We assume Gaussian profiles for pseudo objects 
whose size distribution (measured by SExtractor) 
matches that of the observed LAEs.
The completeness thus computed is $>0.8$ 
for $NB816 < 25.0$, and is $\simeq 0.75$ at $NB816 = 26.0$.
The correction is thus modest in the whole magnitude range 
of the photometric sample.

\begin{figure}
  \hspace{-35pt}
  \FigureFile(120mm,120mm){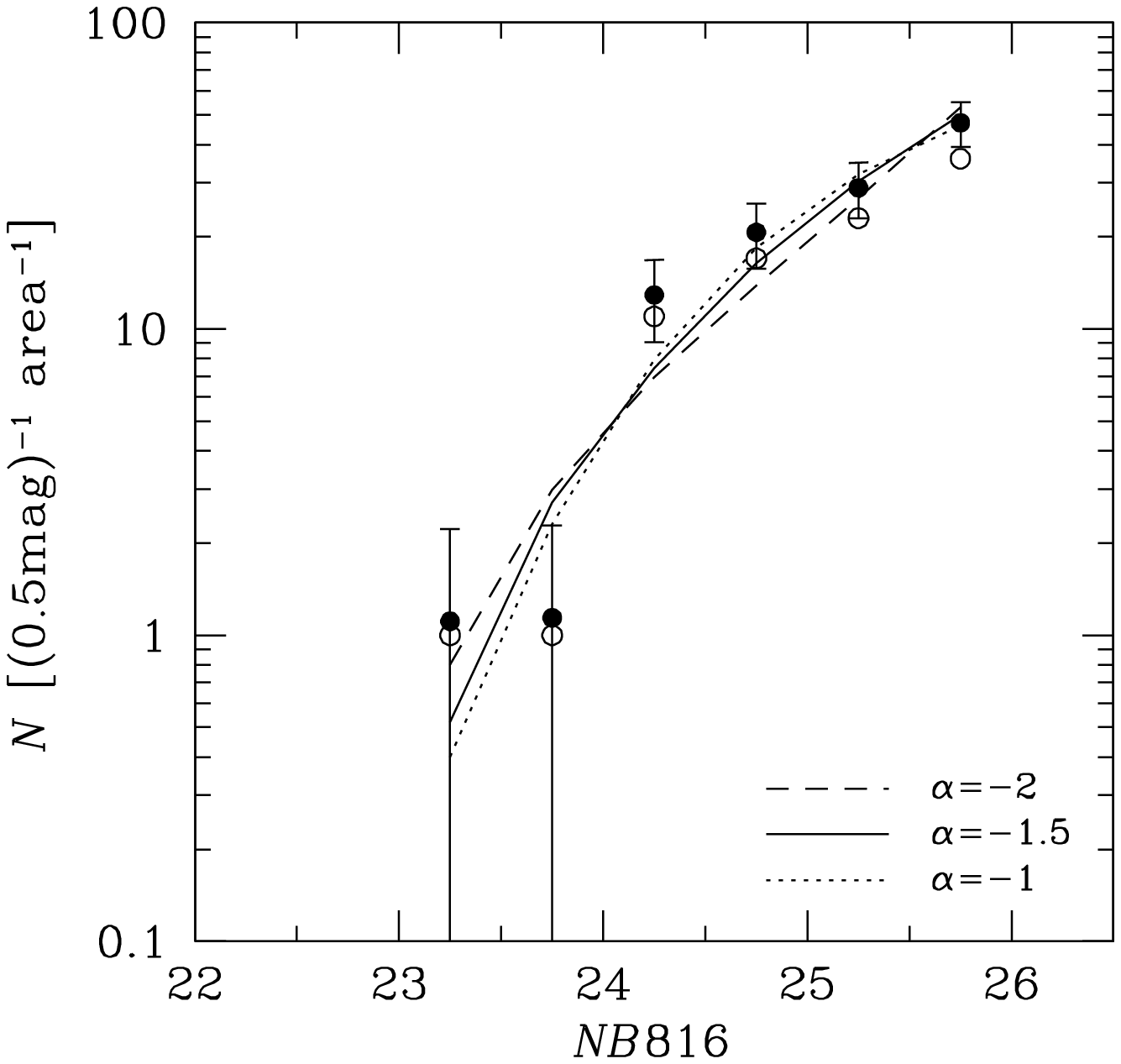}
  \vspace{-30pt}
  \caption{
         Number counts of LAEs in the photometric sample.
         The filled circles represent the counts after 
         correction for detection completeness, while 
         the open circles show the raw counts. 
         The dotted, solid, and dashed lines indicate, respectively, 
         the best-fit Schechter functions 
         for $\alpha=-1, -1.5$, and $-2$ derived in Subsection 5.1.
    \label{fig:nm}}
\end{figure}

We show in Figure \ref{fig:nm} the observed number counts 
by circles and the predicted counts by three curves corresponding to 
the three $\alpha$ values.
The dotted, solid, and dashed lines correspond to 
$\alpha = -1, -1.5$, and $-2$, respectively.
All three models reproduce the observed counts well, although 
shallower $\alpha$ gives a slightly better fit. 
This suggests that the Ly$\alpha$ LF 
of $z=5.7$ LAEs is approximated well by the Schechter function.
The best-fit parameters are 
$(L^\star \hspace{2pt}[{\rm erg}\hspace{2pt}{\rm s}^{-1}], 
\phi^\star \hspace{2pt}[{\rm Mpc}^{-3}])= 
(5.2^{+1.4}_{-1.1} \times 10^{42}, 1.2^{+0.4}_{-0.3} \times 10^{-3}),
(7.9^{+3.0}_{-2.2} \times 10^{42}, 6.3^{+3.0}_{-2.0} \times 10^{-4}),
(1.6^{+0.9}_{-0.6} \times 10^{43}, 1.6^{+1.4}_{-0.7} \times 10^{-4})$
for $\alpha=-1, -1.5, -2.0$, respectively.
The $L^\star$ values are {\lq}observed{\rq} values.  
We assume that half of the intrinsic luminosities are erased 
due to H{\sc i} gas; one can thus double $L^\star$ to obtain intrinsic 
values.

\begin{figure}[h]
  \hspace{-45pt}
  \FigureFile(130mm,130mm){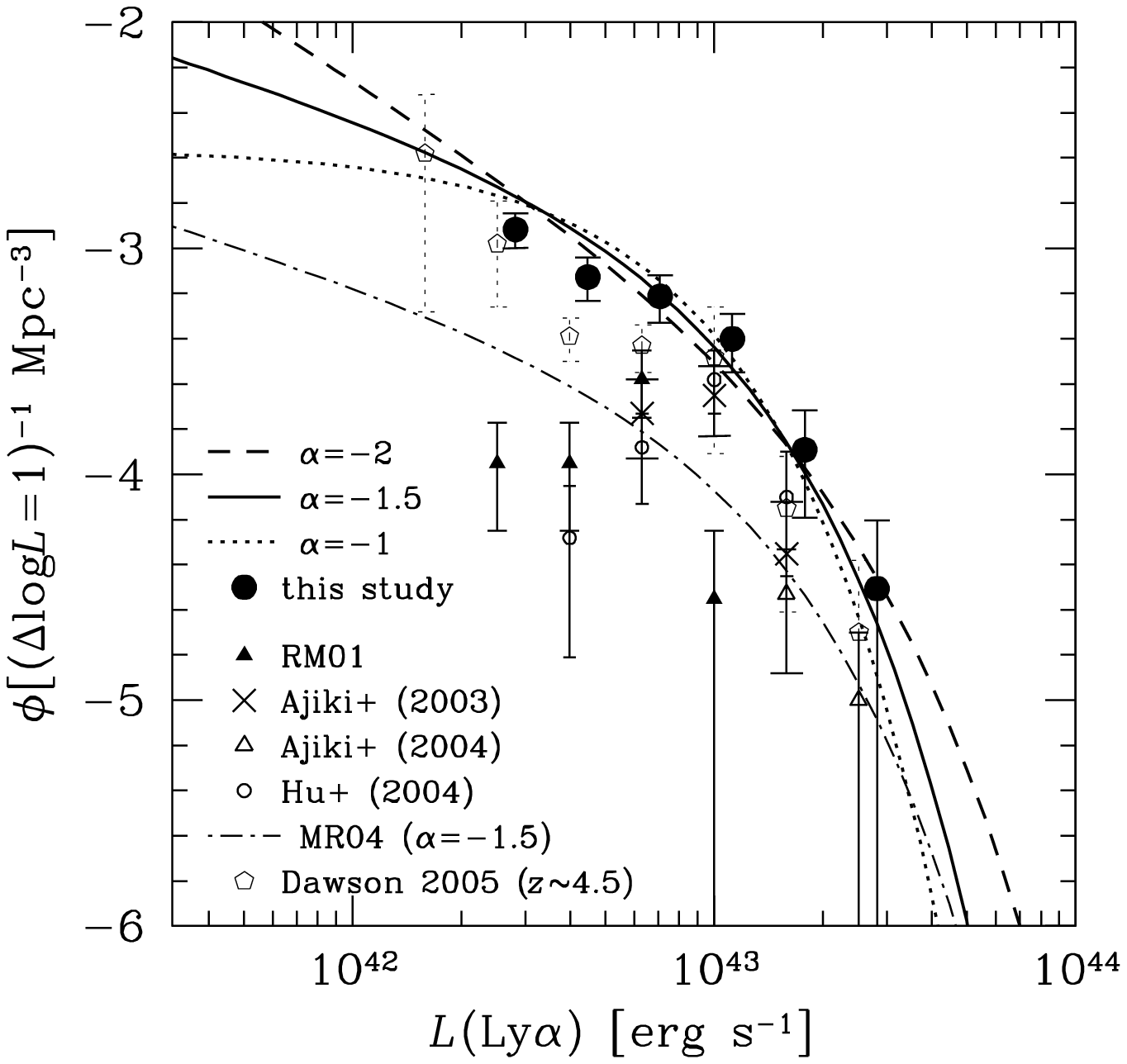}
  \vspace{-25pt}
  \caption{
         Ly$\alpha$ luminosity functions of LAEs at $z=5.7$.
         The dotted, solid, and dashed lines indicate, respectively, 
         the best-fit Schechter functions 
         for $\alpha=-1, -1.5$, and $-2$ derived in Subsection 5.1.
         The filled circles show rough estimates obtained 
         by dividing the number counts of the photometric sample 
         by the effective volume.
         The filled triangles, crosses, open triangles, and 
         open circles indicate the data given by 
         Rhoads \& Malhotra (2001), Ajiki et al. (2003), 
         Ajiki et al. (2004), and Hu et al. (2004).
         The dot-dashed line represents the luminosity function 
         derived in Malhotra \& Rhoads (2004) 
         by fitting the Schechter function to a compiled data set.
         The pentagons represent the LF of LAEs at $z \sim 4.5$ 
         by Dawson (2005).
    \label{fig:lf_LyA}}
\end{figure}

The best-fit Schechter functions obtained above 
for $\alpha=-1, -1.5$, and $-2$ are shown in 
Figure \ref{fig:lf_LyA} 
by the thick dotted, solid, and dashed lines, respectively.
To give some idea about the luminosity range 
in which our Schechter functions are reliable, 
we plot a simple estimate of the luminosity 
function for our sample by the filled circles.
These data points are obtained by dividing the number counts 
in NB816 (after correction for detection completeness) 
by the effective volume.
Here, we calculate the Ly$\alpha$ luminosity of each object 
by subtracting the continuum emission measured from the $z'$-band 
magnitude from the total luminosity in the NB816 band, assuming $z=5.7$.
This is a rough estimate as described above, 
since NB816 magnitude for a given 
set of $L$ and EW varies with redshift.

From Fig.  \ref{fig:lf_LyA}, 
the Schechter functions are found to be reliable 
over $\approx 3 \times 10^{42}$ -- $3 \times 10^{43}$ erg s$^{-1}$. 
In this range, the LFs for all three $\alpha$ values keep 
rising with decreasing $L$, suggesting that there exist many 
faint LAEs beyond our detection limit. 
Indeed, Santos et al. (2004) detected a high number density of 
faint ($L \sim 10^{40-42}$ erg s$^{-1}$) LAEs at 
$z=4.5-5.7$ in a survey of lensed LAEs in very small areas.
Malhotra \& Rhoads (2004) found that a combination of 
Santos et al.'s LF with brighter estimates is consistent 
with Schechter functions of $\alpha \sim -1$ -- $-2$.

Note that the simple estimate of the LF (filled circles) 
agrees fairly well with the best-fit Schechter functions.
This means that the simple method gives a good approximation 
to the true LF at least for our data.
This does not, however, reduce the importance of the accurate 
derivation of the LF made above, 
since we cannot judge the reliability of the simple estimate 
without an accurate calculation.
In addition, the simulations used to derive the LF 
is useful in predicting the redshift distribution of LAEs 
in a flux-limited sample  
as well as placing a constraint on the distribution of EWs.

\subsubsection{Comparison with Previous Measurements}

Overplotted in Figure \ref{fig:lf_LyA} are 
four previous results for $z=5.7$ LAEs, 
which are taken from a compilation by Ajiki et al. (2004).
The filled triangles represent 
the LF given in Rhoads \& Malhotra (2001), 
which is based on a narrow-band sample of 18 candidates 
from a blank field of 710 arcmin$^2$.
The crosses indicate the data in Ajiki et al. (2003) 
from 20 candidates detected 
in a 720 arcmin$^2$ area including 
the QSO SDSSp J104433.04$-$012522.2 at $z=5.74$.
The open triangles show the data in the same QSO field 
but from a survey with an intermediate-band filter (Ajiki et al. 2004).
The open circles indicate Hu et al.'s (2004) measurements
based on spectroscopically confirmed 19 LAEs 
in a blank field of 702 arcmin$^2$. 
The dot-dashed line denotes the LF 
given in Malhotra \& Rhoads (2004) derived 
from a fit of the Schechter function to a compiled data set, 
including Rhoads \& Malhotra (2001), Hu et al. (2004), 
and Ajiki et al. (2004).
All but Ajiki et al. (2003) have not corrected for 
detection completeness.

At $L>1\times 10^{43}$ erg s$^{-1}$, 
our LF is in an acceptable agreement 
with those by Ajiki et al. (2003) and Hu et al. (2004).
Ajiki et al.'s (2004) measurements are lower than ours, 
probably because they adopted a higher threshold for EWs, 
$W{\rm_{0}^{a}} \gtsim 45$ \AA, than the other authors 
including us. 
At $L<1\times 10^{43}$ erg s$^{-1}$, our measurements are 
higher than all of the others by up to an order of magnitude.
We infer that the main reason for this discrepancy is 
that our data are much deeper than the others and 
that the other measurements are not corrected 
for detection completeness.
Large-scale inhomogeneity in the spatial distribution of LAEs 
may be another likely reason for the discrepancy.
The differences in selection criteria for LAEs between 
our sample and the others are probably not a main reason for 
the discrepancy, 
since, for our data, changing the $i'- NB816$ 
threshold over $1.0$ -- $1.5$ does not largely change 
the number of LAEs selected.

The open pentagons in Figure \ref{fig:lf_LyA} represent 
the LF of $z \sim 4.5$ LAEs obtained by Dawson (2005) 
from 76 objects with spectroscopic redshifts.
This LF is found to be slightly {\it lower} than our $z=5.7$ LF, 
but the difference is within the $1 \sigma$ errors 
for most data points.
Thus, drastic evolution in the LAE LF from $z \sim 4.5$ 
to $z=5.7$ is ruled out.
Ouchi et al. (2003) show that the LF at $z=4.9$ 
derived from a sample of 87 LAEs agrees with that for $z=3.4$ LAEs 
by Cowie \& Hu (1998), while Maier et al. (2003) found a significant 
decline in the LAE LF from $z=3.4$ to $z=5.7$ 
in their samples of $z=4.8$ and $z=5.7$.
Our finding is consistent with Ouchi et al.'s (2003).
The decrease found by Maier et al. (2003) could be due to 
the small statistics of their samples ($N=$ 5 and $11$ 
for $z=4.8$ and $5.7$), 
or an incompleteness in object detection in their relatively 
shallow data.

The largest uncertainty inherent in our LF measurements may be 
cosmic variance.
The presence of large-scale structures of LAEs has been reported 
(Shimasaku et al. 2003, 2004; Ouchi et al. 2005).
Hu et al. (2004) also found a highly structured distribution 
of LAEs at $z=5.7$ in their $26.'5 \times 26.'5$ field.
Ouchi et al. (2005) used a catalog of $z = 5.7 \pm 0.05$ LAEs 
for a continuous 1 deg$^2$ area to calculate 
the rms fluctuation of LAE overdensity 
within a sphere of (comoving) 20 Mpc radius to be $40 \pm 20$ \%.
This is the estimate of the cosmic variance of LAEs 
on the largest scale reported so far.
Our survey volume, $1.8 \times 10^5$ Mpc$^3$, 
is five times larger than the volume of a sphere of 20 Mpc radius. 
Thus, the uncertainty in the number density 
of LAEs in our sample is likely to be much smaller than $40\%$, 
unless there exists an unusual inhomogeneity 
in the LAE distribution on scales of $\gg 20$ Mpc.

%
%

\subsection{Equivalent Widths}

\subsubsection{Calculation of Equivalent Widths}

We measure the rest-frame EWs of the spectroscopically confirmed 
34 LAEs by two independent methods.
The first method uses the $z'$-band magnitude to measure 
the continuum flux density, and calculates the EW 
by dividing the continuum-subtracted flux in NB816 
by the flux density of the continuum.
The second calculates the EW and the continuum flux density 
simultaneously from the $i'$ and NB816 magnitudes.
Note that at $z=5.7$ the Ly$\alpha$ line enters 
both $i'$ and NB816 bands.
Malhotra \& Rhoads (2002) used the first method, 
while Hu et al. (2004) applied both and then adopted the first method.
In either method, we do not correct for the effect of H{\sc i} 
absorption to the Ly$\alpha$ emission, 
while taking account of the IGM absorption of 
the continuum emission shortward of the Ly$\alpha$ line 
(Madau 1995).
In what follows, $W{\rm_{0}^{a}}(z')$ and $W{\rm_{0}^{a}}(i')$ 
denote respectively the rest-frame EWs calculated from the first 
and the second methods.

The EW of the majority of our LAEs is highly uncertain 
because of their faint continuum emission; 
uncertainties are in general larger for objects with larger EWs.
Furthermore, the probability distribution function of EW for 
each object, $P(W)dW$, is not symmetric but has a skewed shape 
with a long tail of large EWs, 
since the continuum emission enters into the denominator 
of the expression of EW. 
For each object and each of the above two methods, 
we carefully evaluate $P(W)$ using Monte Carlo simulations 
to calculate the central value of EW and its $1\sigma$ errors 
(see, e.g., Dawson et al. 2004).
In the first method, 
we first assume that each flux density of $NB816$ and $z'$ 
has a Gaussian probability density function 
centered on the measured flux density with a standard deviation 
of the measured error. 
Then we run Monte Carlo simulations to obtain a large set of 
$(NB816, z')$ flux densities. 
Finally, we calculate EWs for individual $(NB816, z')$ combinations 
of the set and derive $P(W)$.
Similar simulations are performed for $NB816$ and $i'$ 
in the second method.

\begin{figure}
  \hspace{-35pt}
  \FigureFile(130mm,130mm){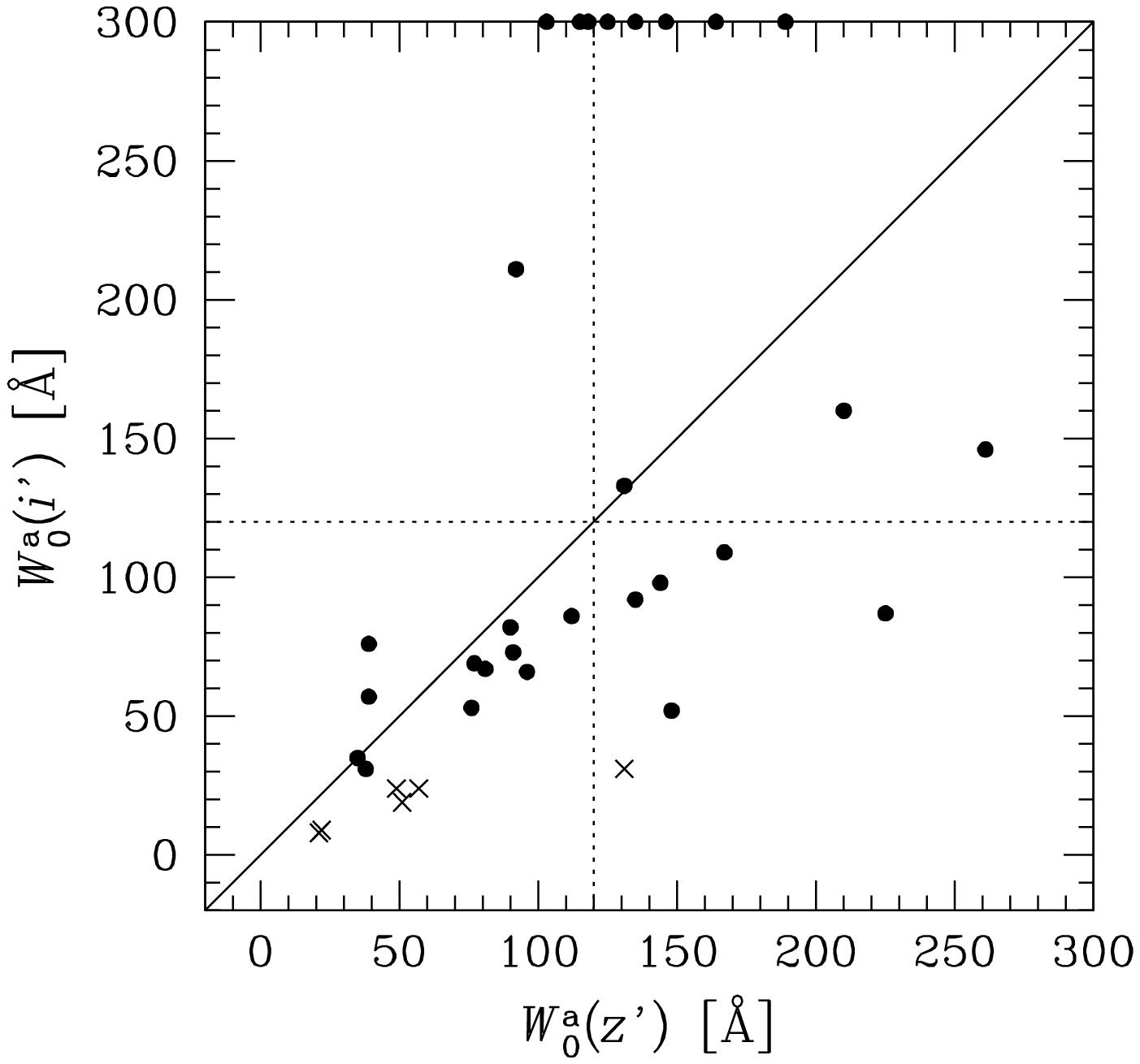}
  \vspace{-30pt}
  \caption{
         $W{\rm_{0}^{a}}(z')$ versus $W{\rm_{0}^{a}}(i')$
         for the 34 LAEs in the spectroscopic sample. 
         The crosses indicate the six objects not meeting 
         the photometric selection criteria for LAEs.
         The dotted lines indicate the maximum allowable value 
         by normal star formation.
    \label{fig:EWi_EWz}}
\end{figure}

The $W{\rm_{0}^{a}}(z')$ and $W{\rm_{0}^{a}}(i')$ values 
of the 34 LAEs derived above are given in Table \ref{tab:LAE_prop}.
Three values in each of the $W{\rm_{0}^{a}}(z')$ and 
$W{\rm_{0}^{a}}(i')$ columns are the central value ($W_c)$, 
$1\sigma$ lower limit ($W_{-}$), 
and $1\sigma$ upper limit ($W_{+}$) from left to right; 
they are defined as $\int_{0}^{W_{c}} P(W)dW = 0.5$ and 
$\int_{W_{-}}^{W_{c}} P(W)dW = \int_{W_{c}}^{W_{+}} P(W)dW = 0.34$.
Figure \ref{fig:EWi_EWz} plots $W{\rm_{0}^{a}}(i')$ against 
$W{\rm_{0}^{a}}(z')$ (both are central values).
The crosses correspond to the six objects with $B$ or $V$ detection; 
they are identical to the six objects not passing 
the photometric selection criteria for LAEs.
For low EW objects $W{\rm_{0}^{a}}(i')$ correlates 
with $W{\rm_{0}^{a}}(z')$, 
but little correlation is seen at $\gtsim 100$ \AA. 
Some objects with $W{\rm_{0}^{a}}(z') \gtsim 100$ \AA\ 
have an extremely large $W{\rm_{0}^{a}}(i')$. 
To derive $W{\rm_{0}^{a}}(i')$, the continuum emission is calculated 
by subtracting the flux falling into the NB816 band from that falling 
into the $i'$ band.
Hence, the continuum emission calculated for objects with 
intrinsically faint continua (large EWs in most cases) 
can be close to zero or negative because of large photometric errors 
in $i'$-band magnitudes. 
Since a negative value is artificial and it actually means 
an extremely large (positive) EW, 
we assign $+ \infty$ to objects with negative $W_c$.

\begin{figure}
  \hspace{-35pt}
  \FigureFile(140mm,140mm){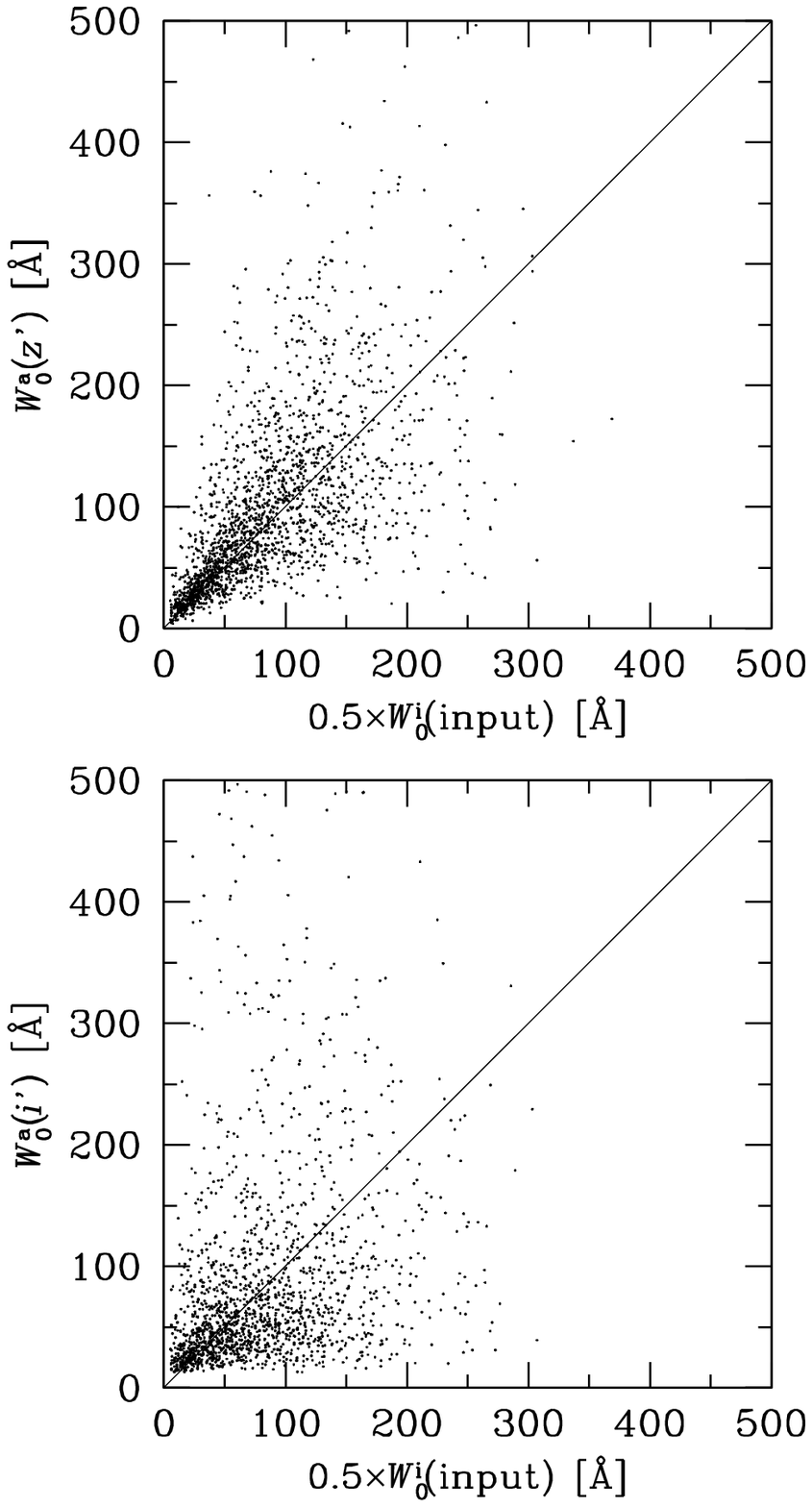}
  \vspace{-30pt}
  \caption{
         Measured equivalent widths plotted against true (input) 
         values for simulated LAEs. 
         {\it Panel (a)}. $W{\rm_{0}^{a}}(z')$ versus 
         $0.5 \times W{\rm_{0}^{i}(input)}$.
         {\it Panel (b)}. $W{\rm_{0}^{a}}(i')$ versus 
         $0.5 \times W{\rm_{0}^{i}(input)}$.
         For both panels 2,000 objects are plotted. 
         They are selected randomly from the simulated catalog 
         with the best-fit Schechter parameters for $\alpha=-1.5$ 
         and $\sigma_W = 200$ \AA. 
         All of the objects satisfy the selection criteria for LAEs.
    \label{fig:EW_simul}}
\end{figure}

\begin{figure}
  \hspace{-45pt}
  \FigureFile(130mm,130mm){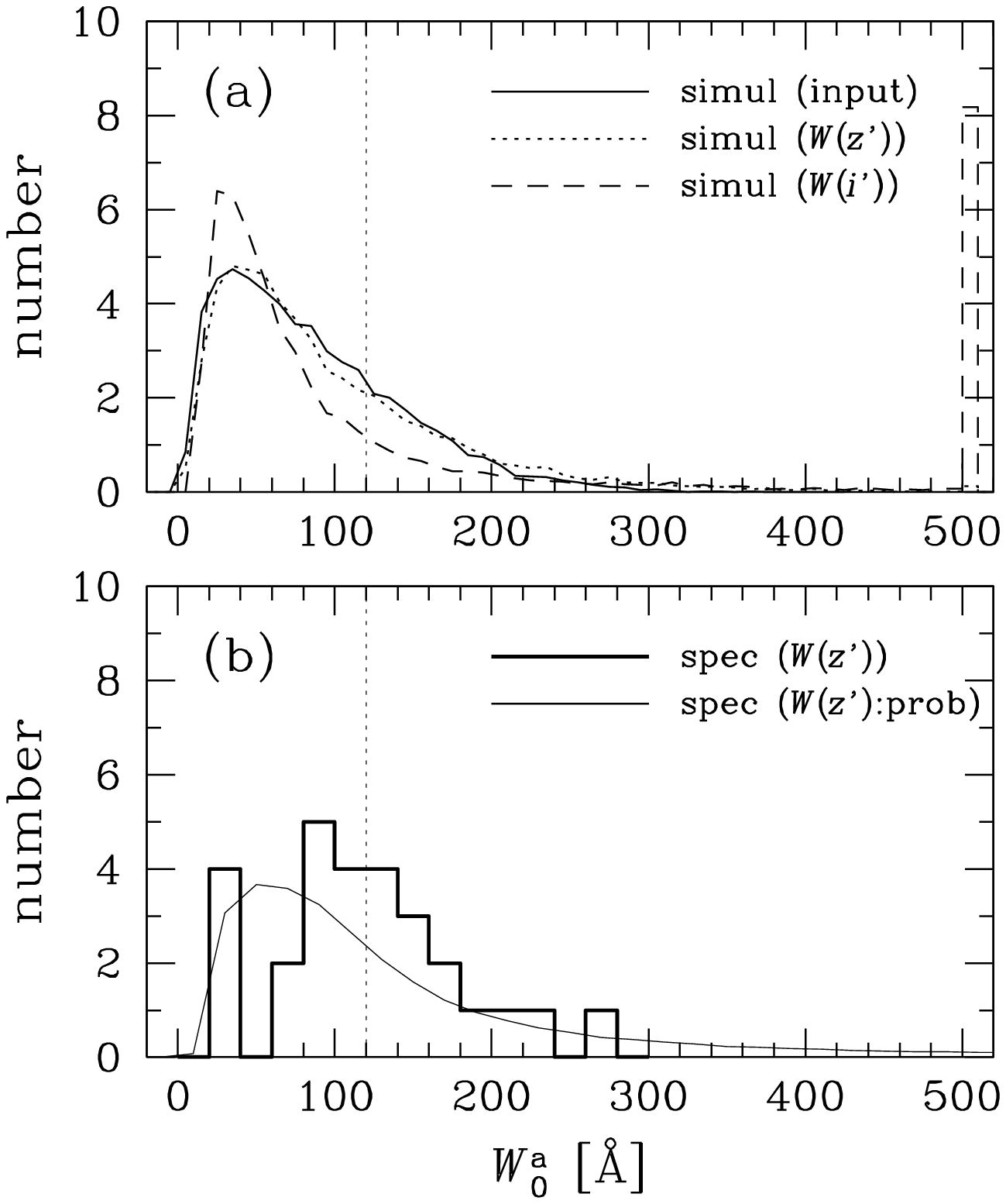}
  \vspace{-45pt}
  \caption{
         {\it Panel (a)}.
         Distributions of input and measured EW values 
         for all simulated LAEs.
         The thin solid curve indicates the input (i.e., true) 
         distribution. 
         The dotted and dashed curves are for 
         $W{\rm_{0}^{a}}(z')$ and $W{\rm_{0}^{a}}(i')$, respectively.
         For $W{\rm_{0}^{a}}(i')$, 
         objects with $W{\rm_{0}^{a}}(i') \ge 500$ \AA\ are 
         shown at $W_0^{\rm a}=505$ \AA.
         {\it Panel (b)}.
         EW distribution of the 28 confirmed LAEs meeting the 
         photometric selection criteria.
         The solid histogram represents the number frequency 
         of $W{\rm_{0}^{a}}(z')$ (central values), while 
         the solid curve corresponds to the sum of their 
         probability distribution functions ($P(W)$).
         In both panels, 
         the dotted line indicates the maximum allowable value 
         by normal star formation.
    \label{fig:EW_hist_simul}}
\end{figure}

The relative robustness against photometric errors between 
the two methods depends on the characteristics of the imaging data. 
We examine the reliability of $W{\rm_{0}^{a}}(z')$ and 
$W{\rm_{0}^{a}}(i')$ measurements using the simulations made
to derive the Ly$\alpha$ LF in Subsection 5.1.
Figure \ref{fig:EW_simul} plots $W{\rm_{0}^{a}}(z')$ (panel [a]) 
and $W{\rm_{0}^{a}}(i')$ (panel [b]) 
against the input value for 2,000 model LAEs randomly selected 
from the simulated objects which satisfy the selection criteria.
The best-fit Schechter parameters for $\alpha=-1.5$ and 
$\sigma_W = 200$ \AA\ are adopted.
It is found that $W{\rm_{0}^{a}}(z')$ shows 
a broad but tighter correlation with the input value 
than $W{\rm_{0}^{a}}(i')$.
Panel (a) of Figure \ref{fig:EW_hist_simul} presents 
the relative number 
distribution of the input and measured EW values for all 
simulated objects satisfying the selection criteria. 
The $W{\rm_{0}^{a}}(z')$ distribution (dotted curve) 
agrees well with the input distribution (thin solid curve), 
while the $W{\rm_{0}^{a}}(i')$ distribution (dashed curve) 
deviates largely in the sense that the number density 
of $W{\rm_{0}^{a}} \sim 100$ -- 200 \AA\ objects is underestimated.
Thus, we conclude that $W{\rm_{0}^{a}}(z')$ is a more reliable 
measurement for our data. 
In what follows, we adopt $W{\rm_{0}^{a}}(z')$ for 
the observed values.

We have inferred that the six objects detected either in $B$ or
$V$ are contaminated by a foreground object.
If this is the case, their equivalent widths derived from
$NB816$ and $z'$ photometry will be lower than the true values
because of overestimated continuum emission
(the effect on Ly$\alpha$ luminosity is little, as $NB816$ and
$z'$ flux densities are similarly contaminated).
This underestimation of EWs is, however, expected to be small,
since (i) among the six objects three are detected in $z'$ 
with both $B-z'>1$ and $V-z'>1$ 
and (ii) three are undetected in $z'$ but two of them 
have $B$ and $V$ magnitudes very close to the detection limits.
In addition, we do not use these six objects
in the discussion of the EW distribution in the next subsection.
Thus, we have not corrected their EW values for foreground
contamination.

\subsubsection{Equivalent Width Distribution}

\begin{figure}
  \hspace{-35pt}
  \FigureFile(120mm,120mm){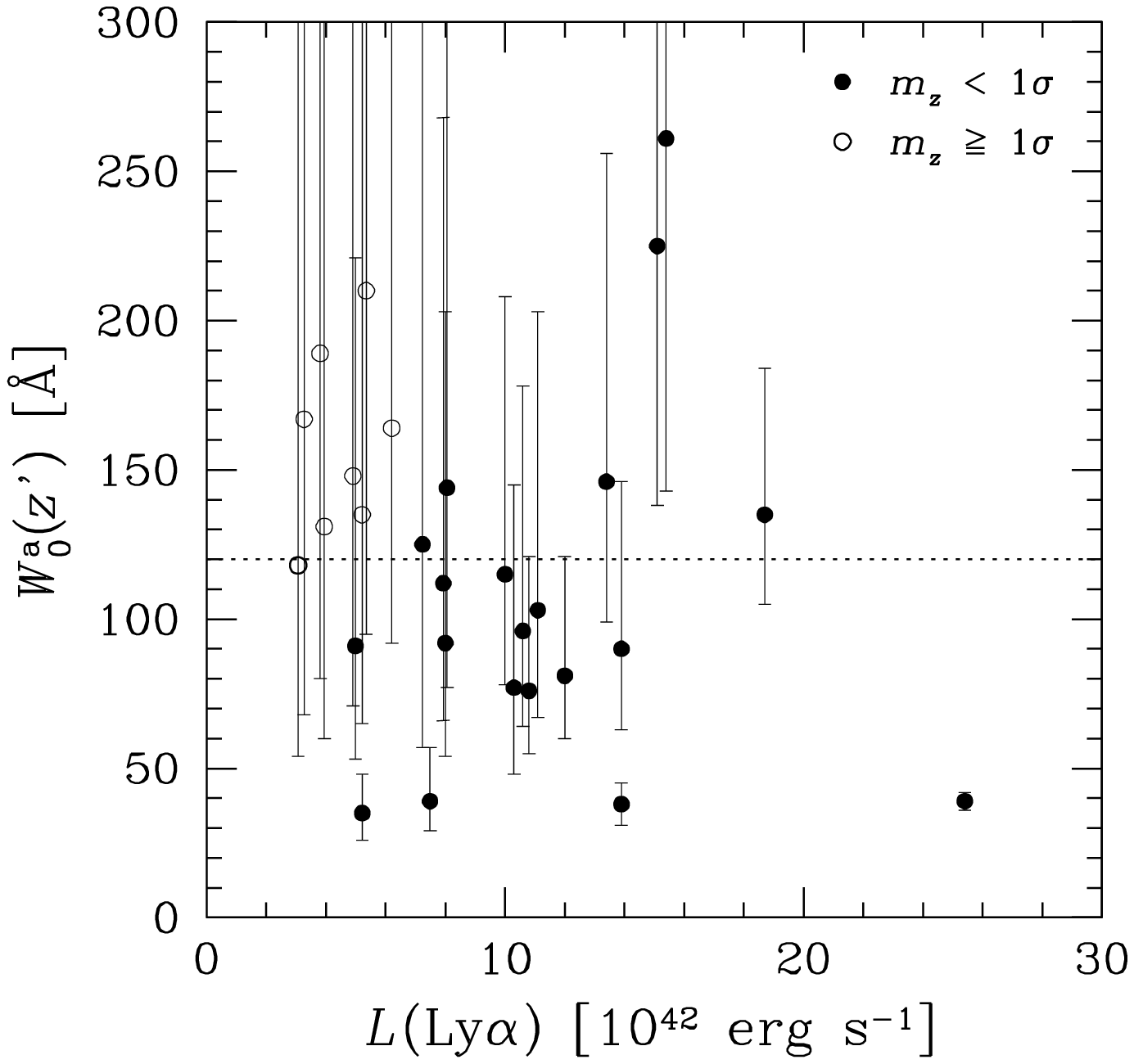}
  \vspace{-20pt}
  \caption{
         $W{\rm_{0}^{a}}(z')$ plotted against Ly$\alpha$ 
         luminosity for the 28 confirmed LAEs which meet 
         the photometric selection criteria.
         Objects shown by open circles have $z'$ magnitudes fainter 
         than the $1\sigma$ magnitude.
         The dotted line indicates the maximum allowable value 
         by normal star formation.
    \label{fig:EW_LLyA}}
\end{figure}

Figure \ref{fig:EW_LLyA} plots $W{\rm_{0}^{a}}(z')$ 
against $L({\rm Ly}\alpha)$ for the 28 confirmed LAEs 
satisfying the photometric selection criteria. 
Objects shown by open circles have $z'$ magnitudes fainter 
than the $1\sigma$ magnitude. 
The horizontal dotted line in Fig. \ref{fig:EW_LLyA} corresponds to 
a maximum value allowed by reasonable stellar populations 
($W{\rm_{0}^{i}} = 240$ \AA) derived by Malhotra \& Rhoads 
(2002). 
Continuously star-forming populations with population-I or -II 
metallicities assuming a Salpeter mass function 
over the normal mass range 
cannot have EWs exceeding this value, 
unless they are as young as $\ll 1 \times 10^7$ yr 
(Charlot \& Fall 1993; Malhotra \& Rhoads 2002).
The number frequency of $W{\rm_{0}^{a}}(z')$ for the 28 LAEs 
is plotted in Figure \ref{fig:EW_hist_simul} (b) by 
the solid histogram.
The solid curve in Fig. \ref{fig:EW_hist_simul} (b) 
corresponds to the sum of the probability distribution functions 
of the 28 objects.

Fig. \ref{fig:EW_LLyA} (and Fig. \ref{fig:EW_hist_simul} [b])
shows that LAEs at $z=5.7$ have very large EWs. 
The median $W{\rm_{0}^{a}}(z')$ among the 28 objects is 
117 \AA\ (or $W{\rm_{0}^{i}}(z') = 233$ \AA), 
with lower and upper quartiles of 86 \AA\ and 147 \AA\ 
(171 \AA\ and 294 \AA\ in $W{\rm_{0}^{i}}$), respectively.
If the probability distribution function (solid curve 
in Fig. \ref{fig:EW_hist_simul} (b)) is used, 
the median, lower quartile, and upper quartile are found 
to be $W{\rm_{0}^{i}}(z') = 213, 125, 386$ \AA, respectively.
The EW distribution based on the spectroscopic sample will 
be biased toward large values, 
since the spectroscopic sample appears to be 
biased toward red $i'- NB816$ colors (i.e., large EWs).
As an independent check, we calculate the median and quartiles 
from the best-fit simulations with $\sigma_W=200$ \AA, 
to find $W{\rm_{0}^{i}} = 147$ \AA\ (median), 
80 \AA\ (lower quartile), and 238 \AA\ (upper quartile).
Although smaller than the corresponding values from 
the spectroscopic sample, these values are still quite large.

\begin{figure}
  \hspace{-45pt}
  \FigureFile(120mm,120mm){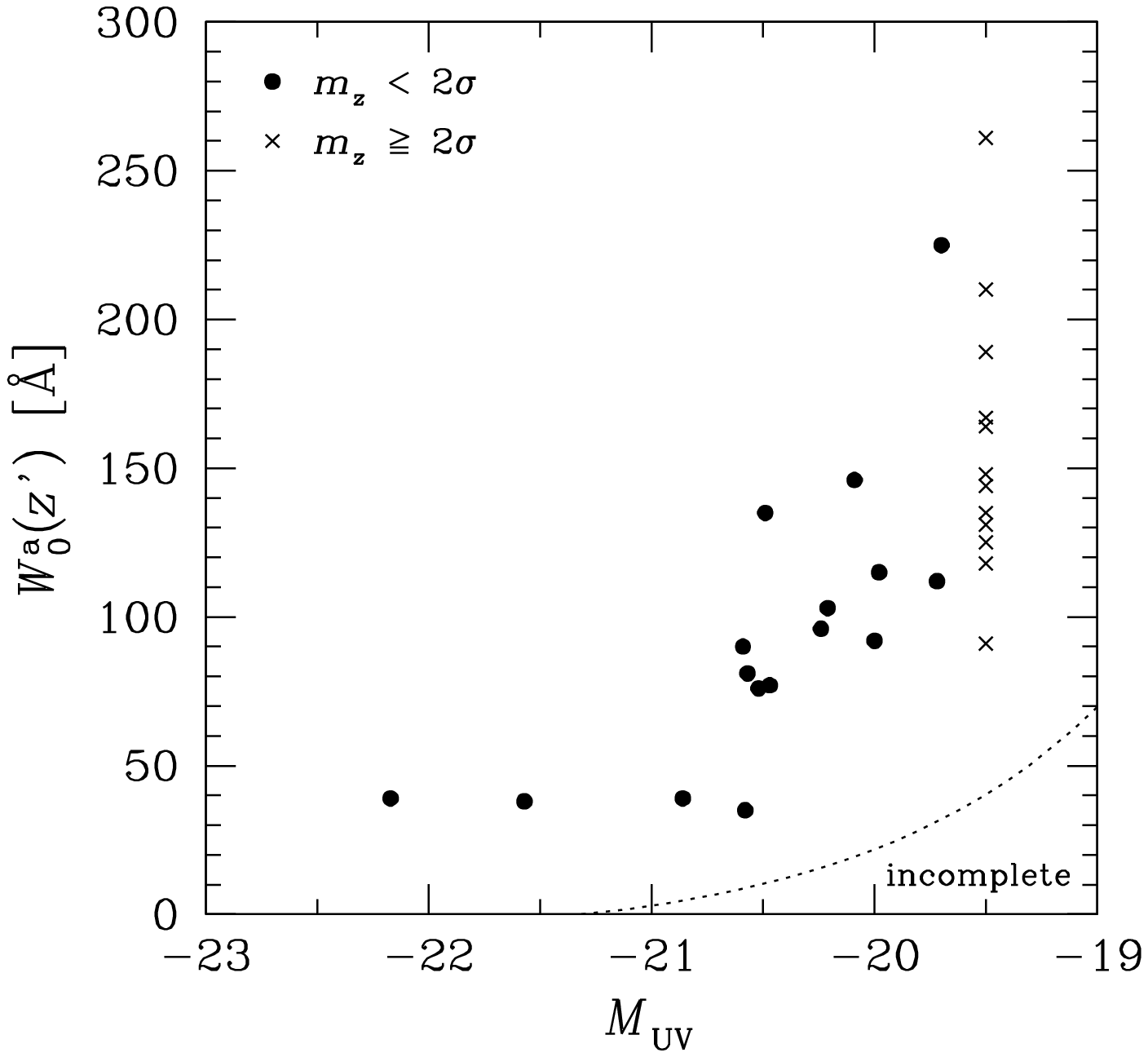}
  \vspace{-20pt}
  \caption{
         Same as Fig. \ref{fig:EW_LLyA}, but plotted 
         against the far ultraviolet absolute magnitude, 
         $M_{\rm UV}$.
         See Subsection 5.3 for the calculation of $M_{\rm UV}$.
         Objects fainter than the $2\sigma$ magnitude in $z'$ 
         are placed at $M_{\rm UV}=-19.5$ and shown by crosses.
         The dotted curve indicates the lower limit of 
         $W{\rm_{0}^{a}}(z')$ detected in our sample 
         ($z=5.7$ is assumed for simplicity); 
         objects below this curve are fainter than $NB816=26.0$ and 
         thus not included in our sample.
    \label{fig:EW_MUV}}
\end{figure}

\begin{figure}
  \hspace{-35pt}
  \FigureFile(120mm,120mm){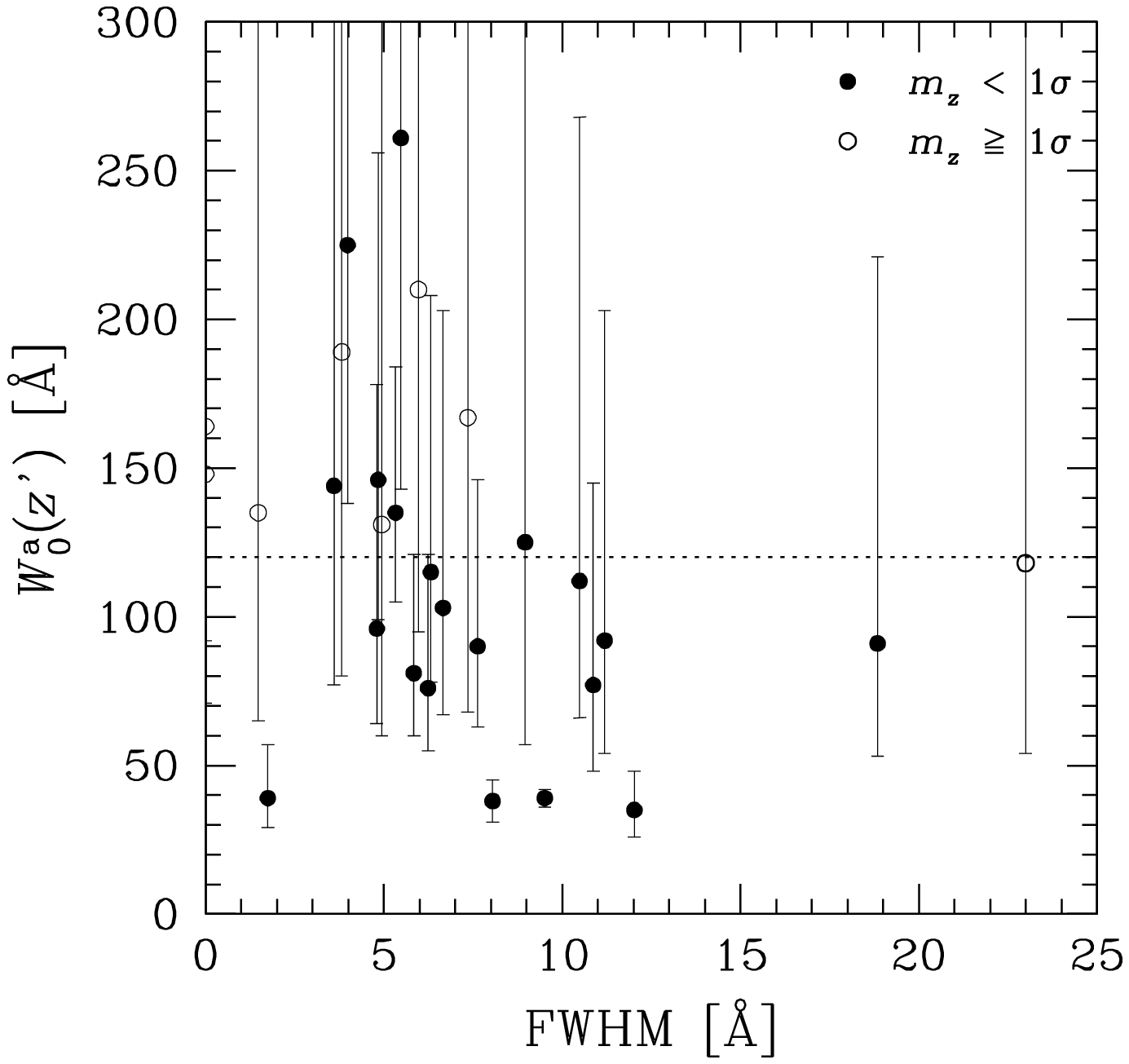}
  \caption{
         Same as Fig. \ref{fig:EW_LLyA}, but plotted 
         against the FWHM of the Ly$\alpha$ line. 
         The FWHM values have been corrected for spectral resolution.
         The dotted line indicates the maximum allowable value 
         by normal star formation.
    \label{fig:EW_FWHM}}
\end{figure}

We do not find a significant correlation 
in $W{\rm_{0}^{a}}(z')$ vs $L({\rm Ly}\alpha)$ 
(Fig. \ref{fig:EW_LLyA}). 
However, we find that objects with fainter far-UV continuum 
luminosities tend to have larger EWs.
Figure \ref{fig:EW_MUV} plots $W{\rm_{0}^{a}}(z')$ 
against far-UV absolute magnitude, $M_{\rm UV}$, 
for the spectroscopic sample. 
See Subsection 5.3 for the calculation of $M_{\rm UV}$.
The dotted curve indicates the lower limit of $W{\rm_{0}^{a}}(z')$ 
detected in our sample; objects below this curve are 
fainter than $NB816=26.0$ and thus not included in our sample.
It is found that all objects brighter than $M_{\rm UV} = -20.7$ 
have $W{\rm_{0}^{a}}(z') \le 50$ \AA. 
On the other hand, the majority of objects fainter than 
$M_{\rm UV} = -20$ have $W{\rm_{0}^{a}}(z') \ge 100$ \AA\ 
unless a large number of objects exist below the dotted curve.
If we interpret that objects with brighter $M_{\rm UV}$ are more 
massive, then the trend seen in Fig. \ref{fig:EW_MUV} 
implies that less massive objects have larger EWs.
Recently, Ando et al. (2005) obtained 11 spectra of $z \sim 5$ LBGs 
with $M_{\rm UV} \ltsim -20.8$, 
finding that bright ($M \ltsim -21.5$) LBGs have EWs less than 
$W{\rm_{0}^{a}} \simeq 20$ \AA\ while the faintest two have large 
EWs of $W{\rm_{0}^{a}} \simeq 40$ -- 80 \AA.
This finding is very close to the trend seen in 
Fig. \ref{fig:EW_MUV}.
It is also interesting to note that a similar trend has been found 
for LBGs at $z \sim 3$ although their absolute EW values 
are much lower than our LAEs' (Shapley et al. 2003).
We also find a weak, negative correlation 
between $W{\rm_{0}^{a}}(z')$ and the FWHM of the Ly$\alpha$ line 
(Fig. \ref{fig:EW_FWHM}). 
This is qualitatively consistent with the trend seen 
in Fig. \ref{fig:EW_MUV}, if the line width is 
due to Doppler broadening.

If one wants to reproduce the observed large median value
by an ordinary, metal-enriched star-forming population with 
a Salpeter IMF (i.e., $x=2.35$) of an upper mass cut of 
$120 M_\odot$, 
one must assume very young ages ($< 1 \times 10^7$ yr), 
negligible dust extinction, and 
a nearly complete absorption of ionizing photons in the system 
(Kudritzki et al. 2000; Malhotra \& Rhoads 2002).
Such young ages mean that we are selectively 
picking up objects which have just started its star formation 
(or just entered a short star-formation phase if they have 
an episodic star-formation history). 
This leads us to an idea that there exist a much larger 
number of ancestors having not started star formation and 
descendants with ages since the onset of star formation too old 
to be selected as LAEs.
This idea appears to be inconsistent with the observed far UV LFs 
of LAEs and LBGs.
In the next section we show that essentially all $z \sim 6$ LBGs 
pass our selection criteria for LAEs. 
This implies that nearly half of LBGs have 
$W{\rm_{0}^{i}} \gtsim 200$ \AA\ and thus younger than 
$1 \times 10^7$ yr.
It looks unlikely that such a high fraction of LBGs fall into 
a so narrow range of age, 
since the LBG selection picks up star-forming galaxies 
irrespective of their ages 
as long as their apparent far UV flux densities are higher
than a given detection limit.
On the other hand, a small portion of our LAEs with equivalent 
widths lower than $\sim 100$ \AA\ can be accounted for by
ordinary Salpeter populations of reasonably old ages 
as long as dust extinction is negligible.

Other possibilities to account for the large median EW are 
top-heavy IMFs or zero-metallicity populations.
Malhotra \& Rhoads (2002) show that both an $x=0.5$ IMF of $1/20$ 
solar metallicity and a Salpeter IMF of zero metallicity can produce 
EWs $\gtsim 200$ \AA\ even if they are $10^8$ yr or older, 
more than ten times older than the ages constrained for 
ordinary Salpeter populations.
A special configuration of the interstellar medium (ISM) may also 
boost the equivalent width.
Hansen \& Oh (2005) argue that Ly$\alpha$ photons can 
preferentially escape from a dusty, multi-phase ISM 
than continuum photons if most of the dust is in cold neutral clouds, 
and thus that the EW of the transmitted spectrum
can be larger than the EW of the unprocessed spectrum 
(see also Neufeld 1991).
If this is the case, the observed large EWs can be reproduced 
by older (or steeper IMF slope) populations.

AGNs can also produce extremely large EWs.
It is, however, unlikely that AGNs dominate in our sample, 
since our LAEs have very narrow rest-frame widths 
of the Ly$\alpha$ emission line, 
with a mean and rms of 1.07 \AA\ and 0.68 \AA, 
or $264$ km s$^{-1}$ and 168 km s$^{-1}$ in velocity, 
after correction for spectral resolution.
Recently, Wang et al. (2004) reported that 
no evidence of AGN activity (including low luminosity quasars) 
has been found for LAEs at $z\sim 4.5$ obtained from a narrow-band 
survey, either in deep X-ray data from {\it Chandra} 
or in optical spectra from Keck 
(see also Malhotra et al. 2003). 
Their report is consistent with our speculation, 
as we are probably selecting a similar population at $z=5.7$ 
to their LAEs.

\subsubsection{Objects Exceeding $W{\rm_{0}^{i}} = 240$ \AA} 

Objects with $W{\rm_{0}^{i}} \ge 240$ \AA\ are particularly 
interesting, since even a population of a very flat IMF of $x=0.5$ 
(but metal enriched) cannot produce such large EWs unless 
it is younger than $1 \times 10^7$ yr (Malhotra \& Rhoads 2002).
We find in the confirmed 28 LAEs that 46 \% (13 objects) have 
$W{\rm_{0}^{i}}(z') \ge 240$ \AA. 
Similarly, the fraction calculated from the probability 
distribution functions of EW is 44 \%.
We can also estimate the fraction from the simulations performed 
in subsection 5.1.1.
The simulations for the best-fit Schechter parameters of 
the Ly$\alpha$ LF (for $\alpha=-1.5$ and $\sigma_W=200$ \AA)  
show that $24\%$ of the objects satisfying 
the selection criteria have input values of 
$W{\rm_{0}^{i}} \ge 240$ \AA. 
Similarly, $29\%$ of the objects satisfying the selection criteria 
have {\lq}measured{\rq} values of $W{\rm_{0}^{i}}(z') \ge 240$ \AA. 
These values are lower than the observed values, 46 \% and 44 \%, 
probably reflecting the bias in $i'- NB816$ color 
in the spectroscopic target selection.
Alternatively, the simulations could underestimate the true fraction, 
since we have not been able to constrain $\sigma_W$ so strongly.

From these results, we infer that the true fraction 
of $W{\rm_{0}^{i}} \ge 240$ \AA\ is $30\%$ -- $40\%$. 
This value is much lower than that obtained by Malhotra \& Rhoads 
(2002), $60\%$, from a photometric sample of $z=4.5$ LAEs 
detected in their narrow-band survey.
Our result may be closer to Hu et al.'s (2004); 
they found in their $z=5.7$ LAE sample that the majority 
(11 out of 15 objects) have EWs less than 240 \AA, 
although they did not clearly present the fraction.
The reason for the difference between our fraction and 
Malhotra \& Rhoads' is not clear. 
The evolutionary effect is unlikely, since it implies that 
the fraction of large-EW objects declines with redshift, 
which is opposite to the increasing fraction of strong Ly$\alpha$ 
emitters in LBGs (see the next subsection).

Recently, Dawson et al. (2004) obtained a more reliable fraction 
of large-EW objects for $z\sim 4.5$ LAEs 
from a sample of 17 objects with spectroscopic confirmation, 
but they calculated the fraction of 
$W{\rm_{0}^{a}} > 240$ \AA\ in our definition, not 
$W{\rm_{0}^{i}} > 240$ \AA.
Their calculation may be regarded as a conservative lower limit, 
since Ly$\alpha$ emission is assumed to be absorption free.
Using the probability distribution function of EW 
from Monte Carlo simulations, 
they found with 90 \% confidence that 18 -- 29 \% of LAEs 
in their sample exceed $240$ \AA.
A similar calculation for our 28 spectroscopic LAEs 
finds that 7 -- 25 \% exceed 
$W{\rm_{0}^{a}}(z') = 240$ \AA\ with 90 \% confidence. 
Thus, even without absorption correction, 
a certain fraction of $z=5.7$ LAEs 
cannot be accounted for by normal star formation; 
the fraction is similar to that found for $z \sim 4.5$ LAEs.
It is worth noting that in our sample we do not see 
evidence of large-EW objects being free from absorption.
Among the six objects with $W{\rm_{0}^{a}} > 150$ \AA, 
at least five have a Ly$\alpha$ profile which shows clear asymmetry 
with a sharp cutoff in the blue side, suggesting a large amount 
of absorption. 

%
%

\subsection{Far-UV Luminosity Function}

\subsubsection{Calculation of the Luminosity Function}

We derive the LF of the far UV continuum 
emission using the photometric sample of 89 LAE candidates.
We transform the $z'$-band magnitude into the far UV continuum 
at the rest-frame 1350 \AA\ assuming that all of the LAEs are 
located at $z=5.7$.
For objects at $z=5.7$, the $z'$ band covers rest-frame 
wavelengths of 1250 \AA\ -- 1480 \AA\ with a weighted center 
of 1350 \AA.
Then the far UV LF is calculated by dividing the number of 
LAEs in each 0.5 magnitude bin by the effective volume 
corresponding to the FWHM of the NB816 bandpass 
($1.80 \times 10^5$ Mpc$^3$).
In this calculation, the detection completeness in NB816 
is taken into account.
We do not convert the Ly$\alpha$ LF obtained in 
Subsection 5.1 into the far UV LF, 
since the conversion is sensitive to the EW distribution, 
which is still uncertain for our purpose here.
For reference, the $M_{\rm UV}$ values of the 34 LAEs 
with spectroscopic redshifts are presented 
in Table \ref{tab:LAE_prop}.

\begin{figure}
  \hspace{-45pt}
  \FigureFile(130mm,130mm){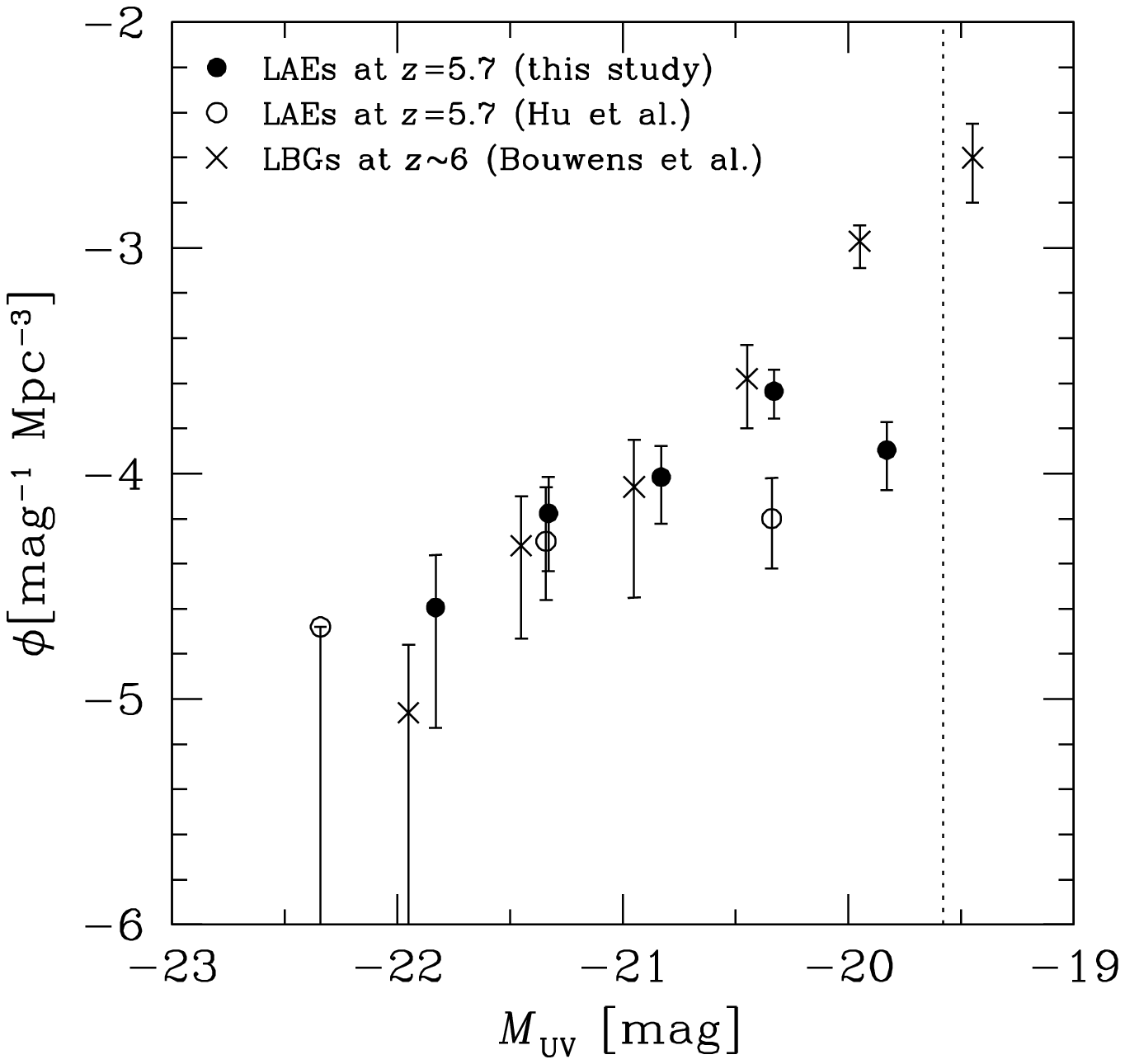}
  \vspace{-30pt}
  \caption{
         Far ultraviolet ($\simeq 1350$ \AA) luminosity functions 
         of LAEs at $z=5.7$ and LBGs at $z \sim 6$.
         The filled circles represent our measurements from 
         the photometric sample.
         The vertical dotted line at $M_{\rm UV}=-19.58$ corresponds 
         to the $2\sigma$ limiting magnitude in the $z'$ band; 
         objects fainter than this magnitude have not been used.
         The open circles show the measurements by Hu et al. (2004).
         The crosses indicate the measurements 
         for LBGs at $z \sim 6$ given in Bouwens et al. (2005).
    \label{fig:lf_FUV}}
\end{figure}

Figure \ref{fig:lf_FUV} plots the far UV LF obtained 
from our photometric sample by filled circles. 
The vertical dotted line at $M_{\rm UV}=-19.58$ corresponds to 
the $2\sigma$ limiting magnitude in the $z'$ band, 27.04 mag; 
objects fainter than this magnitude have not been used 
to derive the LF.
Note that our sample is not $z'$-magnitude limited 
but NB816-band limited ($NB816 \le 26.0$). 
Since the scatter between $z'$ and NB816 magnitudes is large, 
the detection completeness in terms of $z'$ magnitude 
(or far UV magnitude) gradually drops when $z'$ magnitude 
goes fainter because of undetection of objects with relatively 
fainter NB816 magnitudes (for a given $z'$ magnitude).
We find, from the plot of apparent $z'$ magnitude versus NB816 
magnitude, that the measurement of the far UV LF is 
incomplete to some degree at $M_{\rm UV}$ fainter than $\sim -20.5$. 
Therefore, the apparent flattening of the LF at $M_{\rm UV} > -20.5$ 
seen in Figure \ref{fig:lf_FUV} may be 
partly due to this incompleteness.
The open circles show the far UV LF of $z=5.7$ LAEs obtained 
by Hu et al. (2004). 
Their measurements agree with ours very well at $M_{\rm UV}<-21$, 
but their value at $M_{\rm UV}=-20.3$ is lower than ours 
by a factor of four, probably reflecting the detection incompleteness 
of their sample.

\subsubsection{Ly$\alpha$ Emission of LBGs}

We compare the far UV LF of our LAEs with that of $z \sim 6$ LBGs.
Determinations of the far UV LF of $z \sim 6$ LBGs have been 
recently accumulating 
(e.g., Dickinson et al. 2004; Yan \& Windhorst 2004; 
Bunker et al. 2004; Bouwens et al. 2005; Shimasaku et al. 2005); 
see a compilation given in Bouwens et al. (2005).
The differences in $\phi(M)$ among the determinations are 
roughly within a factor of three.
We adopt here Bouwens et al.'s (2005) results 
(crosses in Fig. \ref{fig:lf_FUV}), 
since they are based on a very large sample of 506 LBGs from 
several deep surveys including the GOODS fields and 
the Hubble Ultra Deep Field (HUDF), 
and since their results are in the middle 
among the previous determinations.

LBGs at $z \sim 6$ are detected as $i'$-dropout galaxies 
in multi-color broad-band surveys, and thus they include 
all galaxies with far UV continua brighter than 
a lower limit defined by the survey depth.
In Fig. \ref{fig:lf_FUV}, 
the LAE LF is found to be comparable to the LBG LF 
over the whole magnitude range examined, 
except for our data point at the faintest magnitude.
As stated above, the data point at the faintest magnitude probably 
suffers from a large incompleteness, and thus the true difference 
from the LBG LF will be smaller.
We can thus conclude that at $z \sim 6$ nearly all 
of far-UV selected galaxies with $M_{\rm UV} \ltsim -20$ 
have a strong Ly$\alpha$ emission with $W{\rm_{0}^{i}} \ge 20$ \AA.
In other words, an LAE survey like ours can pick out essentially 
all galaxies brighter than $M_{\rm UV} \sim -20$ at $z \sim 6$.

Malhotra et al. (2005) compared the far UV LF of $z \sim 6$ LBGs 
in the HUDF with the Ly$\alpha$ LF of 
$z=5.7$ LAEs given in Malhotra \& Rhoads (2002), to find similar 
$\phi^\star$ values (after correction for an overdensity seen in 
the HUDF LBGs). 
They argue that the LBG space density could be consistent 
with that of LAEs, but they could not compare the space densities 
of these two populations in a common far UV absolute magnitude 
range.
We demonstrate in this study that the space densities of LBGs 
and LAEs at $z \sim 6$ are similar at least 
down to $M_{\rm UV} \sim -20$.

Figure \ref{fig:lf_FUV} may suggest that the majority of $z \sim 6$ 
LBGs have dust-free, extremely young ages and/or top-heavy IMFs, 
if we recall that our LAEs have extremely large EWs.
A recent observation of far UV colors of $i'$-dropout 
galaxies (i.e., $z \sim 6$ LBGs) by Stanway et al. (2005) 
shows steep (i.e., blue) UV slopes, which might suggest young 
ages ($\sim 10^7$ yr) or top-heavy IMFs, although their sample 
is limited to faint objects ($M_{\rm UV} > -20$) and thus does 
not overlap with our sample.
On the other hand, {\it Spitzer Space Telescope} observations of 
bright ($M_{\rm UV} \ltsim -21$) $i'$-dropout galaxies 
(LBGs at $z \sim 6$) give old ages of a few $\times 10^8$ yr and 
large stellar masses of a few $\times 10^{10} M_\odot$ 
(Yan et al. 2005; Eyles et al. 2005; 
see also Mobasher et al. 2005 for the finding of 
a $J$-dropout galaxy at $z \sim 6.5$ with 
an old age and an extremely large stellar mass). 
{\it Spitzer} observations of HCM 6A (spectroscopically confirmed, 
Ly$\alpha$ emitting galaxy at $z=6.56$ behind Abell 370) 
give very young ages of an order of $10^6$ yr 
(Chary, Stern, \& Eisenhardt 2005), 
but this object is fainter than $-21$.

It is also suggested that bright LBGs have small EWs; 
Bunker et al. (2003) and Stanway et al. (2004) made spectroscopy 
of six $z \sim 6$ LBGs with $M_{\rm UV} \ltsim -21$, finding that 
two have $W{\rm_{0}^{a}} = 20$ -- 30 \AA\ and the rest have no 
detectable Ly$\alpha$ emission  
(A similar trend was found for $z \sim 5$ LBGs by Ando et al. 2005).
This supports our finding that LAEs with brighter 
$M_{\rm UV}$ have on average smaller EWs (Fig. \ref{fig:EW_MUV}).
These results collectively suggest that 
there is a large variety in the star formation history 
in $z \sim 6$ galaxies depending on their mass; 
massive galaxies are considerably evolved while low-mass 
galaxies tend to be very young.

Our study reveals that 
LBGs at $z \sim 6$ have much larger EWs than those at $z \sim 3$.
Steidel et al. (2000) and Shapley et al. (2003) found that 
20 -- 25\% of bright ($M_{\rm UV} < -20$) LBGs at $z\sim 3$ have 
$W{\rm_{0}^{i}} \ge 20$ \AA\ 
(Their measurements include Ly$\alpha$ absorption).
This value is a factor 4 -- 5 lower than that obtained above 
for $z \sim 6$ LBGs, since almost all $z \sim 6$ LBGs 
pass our LAE selection, $W{\rm_{0}^{i}} \gtsim 20$ \AA.
Moreover, the fraction of LBGs 
with $W{\rm_{0}^{i}} \ge 100$ \AA\ is $\sim 2\%$ at 
$z \sim 3$ (Shapley et al. 2003), but this fraction 
rises to as high as about $80\%$ at $z \sim 6$, 
since 24 out of the 28 spectroscopic LAEs 
have $W{\rm_{0}^{i}} \ge 100$ \AA\ 
(The probability distribution functions of the 28 LAEs 
give $82\%$).
Our finding of high fractions of large-EW LBGs is qualitatively 
consistent with the recent finding by Nagao et al. (2005); 
they found that at least three out of 48 $i'$-dropout galaxies 
at $z \sim 6$ have $W{\rm_{0}^{i}} \ge 200$ \AA. 

The extremely high fraction of large-EW LBGs at $z \sim 6$ 
suggests strong evolution 
of Ly$\alpha$ properties of far-UV selected star-forming galaxies 
over $z \sim 6$ and $z \sim 3$.
The origin of this evolution cannot be specified in this study, 
but possible candidates will include 
lower dust extinction, younger stellar ages, 
and drastic changes in metallicity and the IMF slope 
toward lower and flatter values, respectively, 
with increasing redshift.
The increase in the fraction of large EW objects with redshift 
seems qualitatively in accord with 
the recent finding that the far UV continuum of LBGs is on the 
average bluer at $z \sim 6$ than at $z \sim 3$ 
(Bouwens et al. 2005; Stanway et al. 2005). 

We calculate the star formation rate for the 89 LAE candidates 
from the far-UV absolute magnitude using 
the formula given in Madau et al. (1998).
This formula has been applied to LBGs and LAEs at various redshifts 
in previous studies.
Our calculation is limited to the objects brighter than 
$M_{\rm UV} = -19.58$.
The star formation rates span $\simeq 4$ -- $40 M_\odot$ yr$^{-1}$. 
By dividing the sum of the star formation rates by the survey volume, 
we obtain the star formation rate density 
to be $2.3 \times 10^{-3} M_\odot$ yr$^{-1}$ Mpc$^{-3}$. 
This value 
is twice as high as the estimate given by Ajiki et al. (2003), 
reflecting the deeper limiting magnitude of our data, 
and thus is regarded as a new lower limit 
of the far-UV based star formation rate density 
in the $z \sim 6$ universe from LAE surveys.
Our value is about one third that derived from 
the LF of $z \sim 6$ LBGs ($i'$-dropout galaxies) 
down to $0.3L^\star_{z=3}$ (or $-19.6$ mag) 
by Bouwens et al. (2005), 
$\simeq 6 \times 10^{-3} M_\odot$ yr$^{-1}$ Mpc$^{-3}$. 

Note, however, that the formula used above to convert $M_{\rm UV}$ 
into star-formation rate, given in Madau et al. (1998), assumes 
a constant star formation of solar metallicity with a Salpeter IMF 
of ages of $\gg 10^7$ yr. 
Thus, this formula cannot be applied to our LAEs 
if they have very young ($<10^7$ yr) ages or top-heavy IMFs 
or zero metallicity as suggested from the large EW values.
The star formation rate will be underestimated by this formula 
for very young populations, while it will be overestimated 
for populations with top-heavy IMFs.
Therefore, while we can robustly conclude that LAEs have 
a significant contribution to the total far-UV luminosity density 
in the $z \sim 6$ universe, 
the star formation rate density itself 
could contain a large systematic error. 
Furthermore, recalling that nearly all LBGs 
at $z \sim 6$ have EWs large enough to be selected as LAEs, 
we suspect that the star-formation rate density measurements 
for LBGs may also suffer from a similar uncertainty.

\subsubsection{Ly$\alpha$ Bias for the LF of $z \sim 6$ LBGs}

Finally, let us discuss an interesting implication of the large EWs 
of $z \sim 6$ LBGs for their far UV LF. 
In most cases, LBGs at $z \sim 6$ were selected from their red 
$i'-z'$ colors, and their $M_{\rm UV}$ magnitudes were calculated 
from $z'$-band photometry 
(An important exception is the work by Shimasaku et al. 2005). 
At $z=6$, however, the redshifted Ly$\alpha$ line is located 
around the peak transmission of the $z'$ band.
In what follows we take the F850LP band of ACS as a typical $z'$ band.
If an LBG at $z \sim 6$ has $W{\rm_{0}^{i}}=200$ \AA, 
then its F850LP-band magnitude 
will be about 0.8 mag brighter than the value expected in the 
case of no Ly$\alpha$ emission, since its observed-frame 
equivalent width will be as large as 1300 \AA\ 
(exact values depending on redshift), which is nearly the same 
as the FWHM of the F850LP band.
This simple calculation implies that the far-UV LF of $z \sim 6$ 
LBGs calculated from $z'$-band magnitudes will be biased 
toward higher luminosities. 
This bias is stronger for fainter LBGs, since fainter LBGs have 
larger EWs on average (Fig. \ref{fig:EW_MUV}).
Comparing the LF at $z \sim 6$ with those at lower redshifts 
will require correction to the former for this bias.
The correction not only will strengthen the recent finding that 
the number density of bright LBGs drops from $z \sim 3$ 
to $z \sim 6$ 
(e.g., Ouchi et al. 2004; Shimasaku et al. 2005; Bouwens et al. 2005), 
but also might find a new trend that faint LBGs also decrease 
in number toward $z \sim 6$.
Nagao et al. (2004, 2005) have also argued significant brightening 
of $z'$-band magnitudes of $i'$-dropout galaxies due to large EWs.

%
%

\subsection{Spatial Distribution}

\begin{figure}
  \hspace{-45pt}
  \FigureFile(130mm,130mm){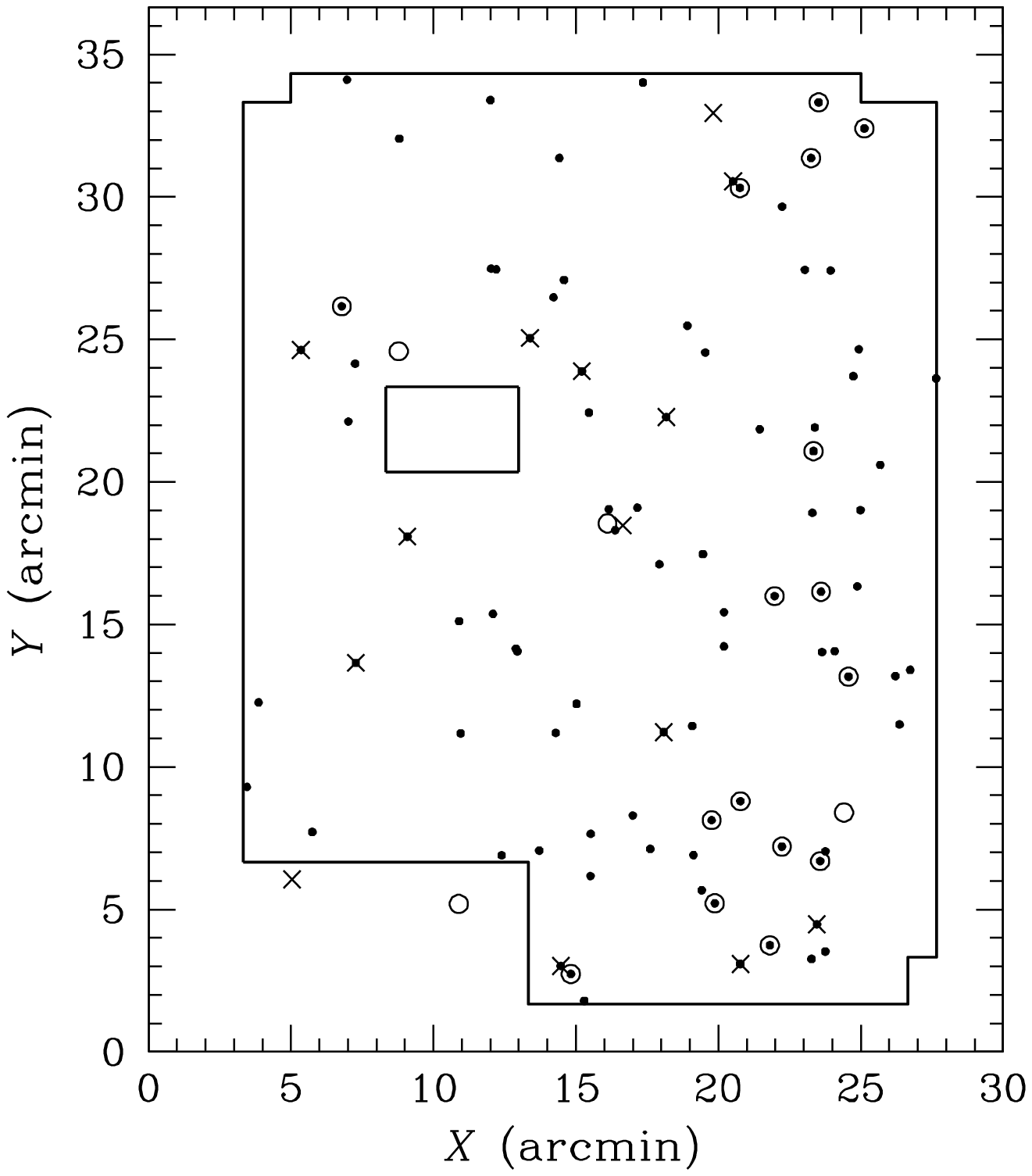}
  \vspace{-30pt}
  \caption{
         Sky distribution of the LAEs in the photometric and 
         spectroscopic samples.
         The small dots indicate the 89 candidates in the photometric 
         sample.
         The open circles and crosses represent, respectively, 
         34 LAEs in the spectroscopic sample with $z \le 5.7$ and 
         $z > 5.7$.
         The thick solid lines outline the region used to make 
         the photometric sample. 
    \label{fig:xy}}
\end{figure}

The sky distribution of our LAEs is shown in Figure \ref{fig:xy}.
The small dots indicate the 89 candidates in the photometric 
sample.
The open circles and crosses represent, respectively, 
34 LAEs in the spectroscopic sample with $z \le 5.7$ and $z > 5.7$. 
The thick solid lines outline the region used to make 
the photometric sample. 
Some spectroscopic objects do not have a counterpart
in the photometric sample because they are either 
out of the region used to construct the photometric sample, 
or ruled out by the selection criteria.

A large-scale density contrast is seen in the sky distribution 
of the 89 candidates. 
The average surface number density over the whole region is 
$0.12 \pm 0.01$ arcmin$^{-2}$.
When the whole region is split into two at $X=17'$, 
the surface number density is $0.097 \pm 0.016$ arcmin$^{-2}$ 
for the eastern half and $0.15 \pm 0.02$ arcmin$^{-2}$ 
for the western half; these values differ from the average 
about $\pm 20\%$.
The line-of-sight distribution of the spectroscopic sample 
also shows inhomogeneity.
Objects with $z \le 5.7$ dominate in the western half, 
while those with $z > 5.7$ dominate in the eastern half.
These findings collectively suggest the existence of 
large-scale structure in our sample.

\begin{figure}
  \hspace{-45pt}
  \FigureFile(130mm,130mm){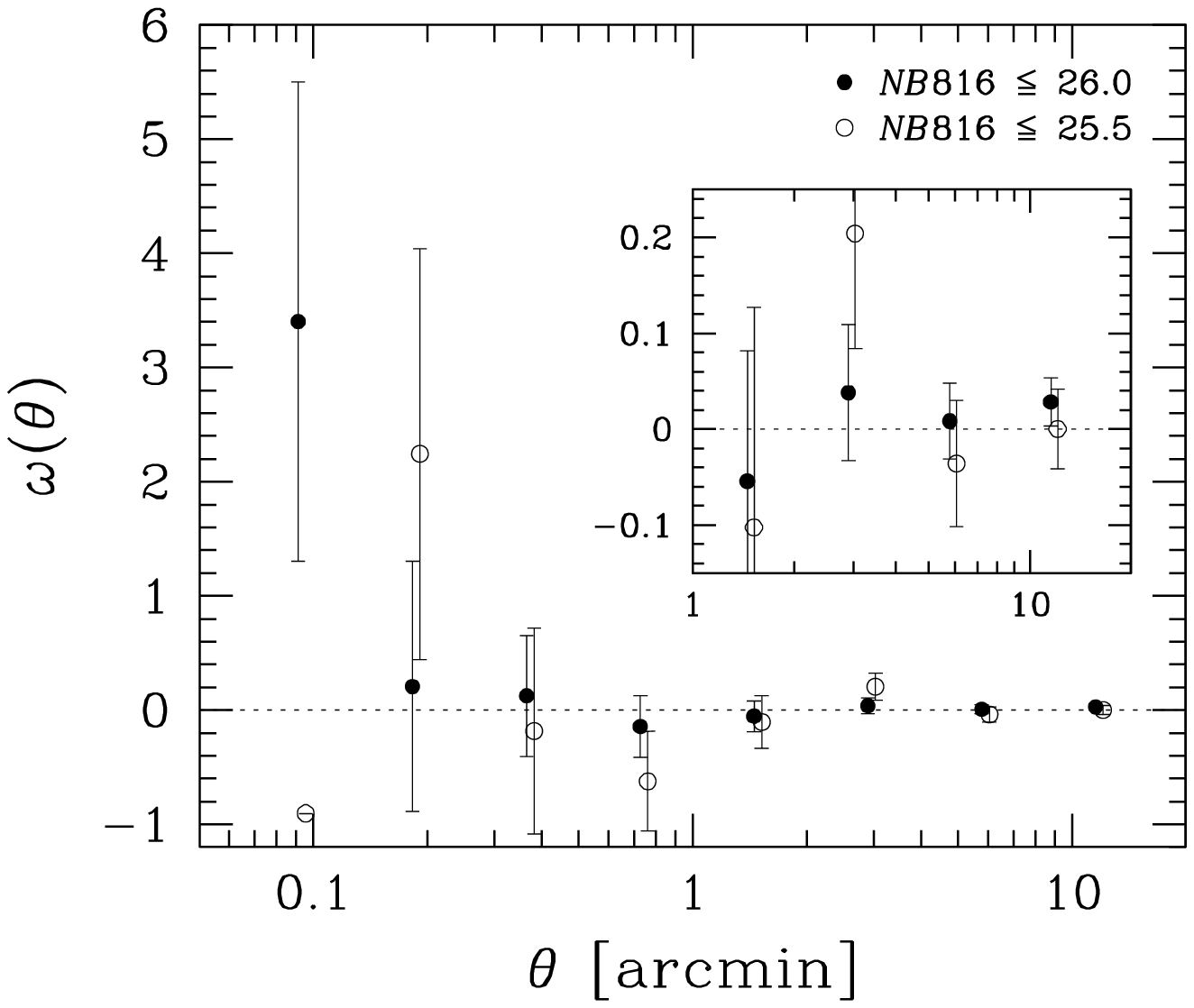}
  \vspace{-45pt}
  \caption{
         Angular correlation function of LAEs in the photometric 
         sample.
         The filled and open circles correspond to the whole sample
         ($N=89$) and a bright ($NB816 \le 25.5$) sample 
         ($N=53$), respectively.
         The inset shows an expanded view at $\theta \ge 1'$.
         The data points have been shifted to the horizontal 
         direction by $-0.01$ and $+0.01$ for the whole and bright 
         samples, respectively, for clarity.
    \label{fig:acf}}
\end{figure}

Figure \ref{fig:acf} plots the angular correlation function 
for the photometric sample (filled circles). 
Clustering signals are found not to be significant; 
the large-scale inhomogeneity found above is not strong 
enough to be clearly reflected in angular correlation.
We also calculate the angular correlation for a bright subsample 
of $NB816 \le 25.5$ (open circles), since brighter galaxies 
are in general known to be clustered more strongly 
at high redshift as well, 
but we do not see significant clustering.
We, however, find possible clustering of the $\simeq 2\sigma$ level 
at a very small separation of $<10''$ in the angular correlation 
function, which might suggest that 
LAEs tend to form close pairs.

Ouchi et al. (2005) have examined the sky distribution of 
$z=5.7$ LAEs in a 1 deg$^2$ area, 
about five times larger than our survey area, and found 
filamentary large-scale structures on scales of several tens of Mpc.
The large-scale structure found in our sample appears 
not to be as prominent as the filamentary structures of 
Ouchi et al. sample.
A similar observation has been obtained for $z\simeq 4.8$ LAEs 
by Shimasaku et al. (2003, 2004).
They have surveyed for LAEs in $z=4.79 \pm 0.04$ and 
$z=4.86 \pm 0.03$ slices of the universe in a $25' \times 45'$ area 
of the SDF, and found a large-scale structure in the latter slice, 
while finding no angular clustering in the former slice.

%
%

\section{Summary}

We have presented the results of a deep survey of 
Ly$\alpha$ emitters (LAEs) at $z=5.7$ 
in the Subaru Deep Field.
A total of 89 LAE candidates with $NB816 \le 26.0$ 
were selected from imaging data of six bandpasses, including 
the narrow band NB816, taken with Suprime-Cam on Subaru.
These candidates have $i'- NB816 \ge 1.5$, red $R-z'$ colors, 
and $B$ and $V$ fluxes fainter than the $2\sigma$ limits.
Spectra of 39 objects satisfying the photometric selection 
criteria for LAEs were obtained with FOCAS on Subaru and 
DEIMOS on Keck II. 
The spectroscopic classification of objects was made 
using weighted skewness, which quantifies 
the asymmetry of the emission line.
Among the 39 objects, 28 are confirmed LAEs, 
one is a nearby galaxy, and eight are unclassified. 
Among the 28 confirmed LAEs, one is outside the region used 
to construct the photometric sample of LAEs.
Also obtained were spectra of another 24 NB816-excess objects 
in the field; six are LAEs and 18 are nearby galaxies. 
We have thus 34 confirmed LAEs in total.
The spectroscopic results show that the photometric sample 
of 89 LAEs is highly reliable with a low contamination.

Using these photometric and spectroscopic samples of LAEs, 
we have studied the LFs, the equivalent width distribution, 
and the spatial distribution of LAEs at $z=5.7$. 
We summarize the main results below.

\begin{itemize}

\item 
By generating mock catalogs of LAEs, 
we have searched for the Schechter parameters of the Ly$\alpha$ 
LF of $z=5.7$ LAEs which best fit 
the observed number counts in the NB816 band.
We have found 
$(L^\star \hspace{2pt}[{\rm erg}\hspace{2pt}{\rm s}^{-1}], 
\phi^\star \hspace{2pt}[{\rm Mpc}^{-3}])= 
(5.2^{+1.4}_{-1.1} \times 10^{42}, 1.2^{+0.4}_{-0.3} \times 10^{-3}),
(7.9^{+3.0}_{-2.2} \times 10^{42}, 6.3^{+3.0}_{-2.0} \times 10^{-4}),
(1.6^{+0.9}_{-0.6} \times 10^{43}, 1.6^{+1.4}_{-0.7} \times 10^{-4})$
for $\alpha=-1, -1.5, -2.0$, respectively, 
over the range where our data are available, 
$L \simeq 3 \times 10^{42}$ -- $3 \times 10^{43}$ erg s$^{-1}$.
The Ly$\alpha$ LF of $z=5.7$ LAEs 
is reproduced well by the Schechter function.
Our LF is consistent with the previous measurements 
at $L>1 \times 10^{43}$ erg s$^{-1}$ within the errors, 
but largely overshoots them at $L<1 \times 10^{43}$ erg s$^{-1}$, 
probably reflecting the detection incompleteness of the previous 
surveys.
No significant evolution is found in the LF between $z \sim 4.5$ 
and $z=5.7$.

\item 
We have measured rest-frame Ly$\alpha$ equivalent widths 
for the spectroscopically confirmed LAEs.
We have found that the median value among the 28 LAEs  
satisfying the photometric selection criteria is 
$W{\rm_{0}^{i}} = 233$ \AA, 
with lower and upper quartiles of 171 \AA\ and 294 \AA, respectively, 
where $W{\rm_{0}^{i}}$, which we regard as the intrinsic value, 
is twice the measured value.
About 30\% -- $40\%$ of LAEs at $z=5.7$ are inferred to exceed 
$W{\rm_{0}^{i}} = 240$ \AA, from both the spectroscopic 
sample and the simulations.
LAEs with $W{\rm_{0}^{i}} \gtsim 200$ \AA\ are unlikely to be 
accounted for by metal enriched, continuously star-forming 
populations with a Salpeter IMF; 
metal-free populations or top-heavy IMFs are required.
We have also found that LAEs with fainter far-UV magnitudes 
have larger EWs.

\item 
We have derived the far-UV ($\simeq 1350$ \AA) LF for LAEs at $z=5.7$ 
down to $M_{\rm UV} \simeq -19.6$ from the photometric sample. 
We have found that this LF agrees with 
the far-UV LF of LBGs at similar redshifts over the whole 
magnitude range where the completeness of our sample 
is expected to be relatively high.
This implies that at $z\sim 6$ almost all 
star-forming galaxies brighter than $M_{\rm UV} \sim -20$ 
have Ly$\alpha$ emission strong enough to be detected as LAEs 
($W{\rm_{0}^{i}} \ge 20$ \AA). 
Moreover, the fraction of $z \sim 6$ LBGs with 
$W{\rm_{0}^{i}} \ge 100$ \AA\ is found to be about 80\%.
These extremely high fractions are in sharp contrast to 
lower-redshift LBGs, suggesting drastic evolution 
in Ly$\alpha$ emission properties.
We have also argued that the far-UV LF and the star formation rates 
of star-forming galaxies at $z \sim 6$ measured 
from $z'$-band photometry could contain large systematic errors.

\item
We have found large-scale structure in the spatial 
distribution of our sample both in the tangential and radial 
directions, although it is not as prominent as the filamentary 
structures seen in Ouchi et al.'s (2005) much larger survey 
for $z=5.7$ LAEs.

\end{itemize}

%
%

\vspace{20pt}
We thank an anonymous referee for valuable comments 
which have much improved the paper.
We are grateful to the staff of the Subaru and Keck Telescopes.
This research would not be made possible without 
the unique and excellent data obtained for  
the Subaru Deep Field Project.
We deeply thank all members of the project who are not co-authors 
of this paper.
N.K. acknowledges support by the Japan Society for the Promotion
of Science through Grant-in-Aid for Scientific Research 16740118.
M.O. has been supported by the Hubble Fellowship program
through grant HF-01176.01-A awarded by the Space Telescope Science
Institute, which is operated by the Association of Universities
for Research in Astronomy, Inc. under NASA contract NAS 5-26555.

%
%

\clearpage


\begin{table*}
  \caption{Spectroscopic properties
    \label{tab:obj_spec}
  }

\tabcolsep 2pt

\begin{center}

(a) LAEs

    \begin{tabular}{rrrrrrrccrccccccccr}
\hline
\hline
\multicolumn{1}{c}{ID} & 
\multicolumn{3}{c}{$\alpha$ (J2000.0)} &
     \multicolumn{3}{c}{$\delta$ (J2000.0)} &
     $z$ & I & 
     \multicolumn{1}{c}{$S_W$} & 
     $NB816$ & $B$ & $V$ & $R$ & $i'$ & $z'$ & $i'-NB816$ & $R-z'$ \\
\multicolumn{1}{c}{(1)} & 
  \multicolumn{3}{c}{(2)} & 
  \multicolumn{3}{c}{(3)} & 
  (4) & (5) & \multicolumn{1}{c}{(6)} & 
  (7) & (8) & (9) & (10) & (11) & (12) & (13) & (14) \\
\hline
  18699 & 13 & 24 & 38.94 & +27 & 13 & 40.9 & 5.645& D &$ 5.03\pm0.82$ & 25.80 & 28.8 & 28.1 & 28.20 & 27.84 & 27.04 & 1.84 &  1.16 \\ 
  20087 & 13 & 24 & 40.52 & +27 & 13 & 57.9 & 5.724& D &$ 8.18\pm0.23$ & 24.00 & 28.8 & 28.1 & 28.20 & 26.48 & 26.39 & 2.22 &  1.81 \\ 
  20495 & 13 & 24 & 11.88 & +27 & 14 &  2.0 & 5.697& D &$ 7.03\pm0.41$ & 24.77 & 28.8 & 28.1 & 28.20 & 27.52 & 27.04 & 2.33 &  1.16 \\ 
  23759 & 13 & 24 &  7.20 & +27 & 14 & 41.6 & 5.687& F &$ 5.81\pm5.16$ & 24.65 & 28.8 & 28.1 & 28.20 & 27.40 & 26.26 & 2.24 &  1.94 \\ 
  27787 & 13 & 23 & 59.71 & +27 & 15 & 26.3 & 5.707& F &$ 6.03\pm2.25$ & 25.23 & 28.8 & 28.1 & 28.20 & 27.84 & 27.04 & 2.31 &  1.16 \\ 
  31858 & 13 & 24 & 56.80 & +27 & 16 &  9.7 & 5.692& D &$ 6.01\pm0.88$ & 25.57 & 28.8 & 28.1 & 28.20 & 27.84 & 27.04 & 2.21 &  1.16 \\ 
  31765 & 13 & 24 & 15.99 & +27 & 16 & 11.1 & 5.691& F &$ 5.90\pm2.52$ & 24.38 & 28.8 & 28.1 & 28.20 & 26.96 & 26.53 & 2.13 &  1.67 \\ 
  36334 & 13 & 25 & 23.41 & +27 & 17 &  1.3 & 5.705& D &$ 3.75\pm0.43$ & 24.62 & 27.6 & 26.9 & 26.47 & 25.93 & 25.34 & 1.17 &  1.13 \\ 
  39849 & 13 & 23 & 59.17 & +27 & 17 & 40.7 & 5.692& F &$ 4.85\pm1.14$ & 24.21 & 28.8 & 28.1 & 28.20 & 27.38 & 26.82 & 2.86 &  1.37 \\ 
  42576 & 13 & 24 &  5.27 & +27 & 18 & 11.6 & 5.684& F &$ 6.30\pm0.98$ & 24.54 & 28.8 & 28.1 & 28.20 & 27.45 & 26.87 & 2.67 &  1.33 \\ 
  46904 & 13 & 24 & 16.47 & +27 & 19 &  7.7 & 5.665& F &$10.30\pm1.65$ & 24.15 & 28.8 & 28.1 & 28.20 & 26.18 & 25.37 & 1.69 &  2.83 \\ 
  48328 & 13 & 23 & 55.36 & +27 & 19 & 23.7 & 5.693& F &$ 7.28\pm4.88$ & 25.80 & 28.7 & 28.1 & 28.20 & 27.49 & 27.04 & 1.63 &  1.16 \\ 
  50215 & 13 & 24 & 11.87 & +27 & 19 & 48.2 & 5.682& F &$ 6.48\pm1.94$ & 25.30 & 28.8 & 28.1 & 28.20 & 27.84 & 27.04 & 2.36 &  1.16 \\ 
  61418 & 13 & 24 & 24.11 & +27 & 22 & 15.1 & 5.748& D &$ 8.00\pm0.01$ & 25.00 & 28.8 & 28.1 & 28.20 & 27.83 & 27.00 & 2.24 &  1.20 \\ 
  70600 & 13 & 23 & 54.60 & +27 & 24 & 12.7 & 5.654& F &$ 8.77\pm0.80$ & 24.81 & 28.8 & 28.1 & 28.20 & 26.70 & 27.04 & 1.72 &  1.16 \\ 
  73078 & 13 & 25 & 13.27 & +27 & 24 & 42.0 & 5.700& D &$11.62\pm2.14$ & 25.79 & 28.8 & 28.1 & 28.20 & 27.84 & 27.04 & 1.95 &  1.16 \\ 
  84305 & 13 & 24 &  6.39 & +27 & 27 &  4.2 & 5.695& F &$ 8.38\pm1.84$ & 24.28 & 28.8 & 28.1 & 28.20 & 26.96 & 26.49 & 2.23 &  1.71 \\ 
  84720 & 13 & 23 & 58.98 & +27 & 27 & 13.4 & 5.682& F &$20.66\pm6.02$ & 24.47 & 28.8 & 28.1 & 28.20 & 27.45 & 26.94 & 2.17 &  1.26 \\ 
  93966 & 13 & 25 &  5.06 & +27 & 29 & 10.3 & 5.721& F &$ 5.36\pm3.74$ & 25.18 & 28.8 & 28.1 & 28.20 & 27.84 & 27.04 & 2.58 &  1.16 \\ 
  95588 & 13 & 24 & 30.63 & +27 & 29 & 34.5 & 5.738& F &$ 8.60\pm4.55$ & 25.41 & 28.3 & 28.1 & 28.05 & 26.93 & 26.79 & 1.21 &  1.26 \\ 
  96007 & 13 & 24 & 33.09 & +27 & 29 & 38.6 & 5.696& F &$ 6.28\pm5.63$ & 25.20 & 28.7 & 28.1 & 28.20 & 27.47 & 27.04 & 1.72 &  1.16 \\ 
 108164 & 13 & 24 &  0.15 & +27 & 32 & 12.1 & 5.672& F &$ 4.13\pm2.23$ & 25.89 & 28.8 & 28.1 & 28.20 & 27.84 & 27.04 & 1.87 &  1.16 \\ 
 113271 & 13 & 24 & 23.70 & +27 & 33 & 24.8 & 5.710& F &$11.82\pm1.01$ & 23.41 & 28.8 & 28.1 & 27.91 & 25.80 & 24.73 & 2.10 &  3.17 \\ 
 121315 & 13 & 24 & 37.19 & +27 & 35 &  2.3 & 5.716& F &$ 3.38\pm2.33$ & 25.37 & 28.8 & 28.1 & 28.20 & 27.77 & 27.04 & 2.04 &  1.16 \\ 
 124530 & 13 & 25 &  6.48 & +27 & 35 & 44.5 & 5.689& D &$ 3.33\pm0.19$ & 24.62 & 27.4 & 27.5 & 27.17 & 27.01 & 27.04 & 1.88 &  0.13 \\ 
 124783 & 13 & 25 & 22.12 & +27 & 35 & 46.8 & 5.720& D &$ 7.66\pm0.80$ & 25.20 & 28.8 & 28.1 & 28.20 & 27.20 & 26.28 & 1.75 &  1.92 \\ 
 126848 & 13 & 24 & 45.48 & +27 & 36 & 12.6 & 5.718& F &$10.49\pm6.73$ & 25.35 & 28.8 & 28.1 & 28.20 & 27.71 & 27.04 & 2.19 &  1.16 \\ 
 132343 & 13 & 25 & 15.60 & +27 & 37 & 19.9 & 5.697& D &$ 6.28\pm0.83$ & 24.78 & 28.8 & 28.1 & 28.20 & 27.61 & 27.04 & 2.38 &  1.16 \\ 
 152586 & 13 & 24 & 11.89 & +27 & 41 & 31.8 & 5.681& F &$ 6.11\pm1.64$ & 24.18 & 28.8 & 28.1 & 28.20 & 27.03 & 26.63 & 2.24 &  1.57 \\ 
 154296 & 13 & 24 & 13.00 & +27 & 41 & 45.8 & 5.715& F &$ 5.01\pm2.76$ & 25.57 & 28.8 & 28.1 & 28.20 & 27.84 & 27.04 & 1.86 &  1.16 \\ 
 158036 & 13 & 24 &  0.51 & +27 & 42 & 35.3 & 5.676& F &$12.68\pm1.53$ & 24.83 & 28.8 & 28.1 & 28.20 & 27.84 & 27.04 & 2.48 &  1.16 \\ 
 163079 & 13 & 23 & 51.96 & +27 & 43 & 38.2 & 5.676& F &$ 4.38\pm1.21$ & 24.54 & 28.8 & 28.1 & 28.20 & 27.16 & 26.87 & 2.11 &  1.33 \\ 
 166310 & 13 & 24 & 16.13 & +27 & 44 & 11.6 & 5.698& F &$10.61\pm1.80$ & 23.93 & 26.1 & 25.9 & 25.83 & 25.43 & 24.84 & 1.19 &  0.99 \\ 
 168127 & 13 & 23 & 59.27 & +27 & 44 & 33.8 & 5.674& F &$ 6.28\pm0.78$ & 24.27 & 28.8 & 28.1 & 28.20 & 27.06 & 27.04 & 2.34 &  1.16 \\ 
\hline
    \end{tabular}


(b) nearby objects

    \begin{tabular}{rrrrrrrccrccccccccr}
\hline
\hline
ID & \multicolumn{3}{c}{$\alpha$ (J2000.0)} &
     \multicolumn{3}{c}{$\delta$ (J2000.0)} &
     $z$ & I & 
     \multicolumn{1}{c}{$S_W$} & 
     $NB816$ & $B$ & $V$ & $R$ & $i'$ & $z'$ & $i'-NB816$ & $R-z'$ \\
\hline
  28247 & 13 & 24 & 11.70 & +27 & 15 & 30.5 &  ...  & F &$ 2.99\pm0.30$& 24.05 & 27.0 & 26.9 & 26.83 & 26.21 & 26.82 & 2.00 &  0.01 \\ 
  29275 & 13 & 25 & 17.89 & +27 & 15 & 46.4 &  ...  & D &$-6.40\pm0.12$& 22.79 & 25.8 & 25.4 & 24.87 & 24.35 & 23.52 & 1.02 &  1.35 \\ 
  31925 & 13 & 25 & 15.61 & +27 & 16 & 11.0 & 1.182 & D &$-8.49\pm0.20$& 24.40 & 26.4 & 26.1 & 25.88 & 25.62 & 25.17 & 1.02 &  0.71 \\ 
  34775 & 13 & 25 &  8.10 & +27 & 16 & 48.1 &  ...  & D &$ 0.14\pm0.19$& 23.57 & 28.1 & 27.4 & 26.60 & 25.55 & 24.52 & 1.18 &  2.08 \\ 
  38133 & 13 & 23 & 56.28 & +27 & 17 & 26.1 & 1.180 & F &$-1.88\pm0.05$& 22.23 & 24.0 & 23.8 & 23.73 & 23.51 & 23.21 & 1.03 &  0.52 \\ 
  42561 & 13 & 24 &  3.41 & +27 & 18 & 17.8 & 0.630 & F &$ 0.65\pm0.01$& 21.83 & 24.1 & 23.8 & 23.41 & 23.14 & 23.22 & 1.02 &  0.19 \\ 
  56181 & 13 & 25 & 25.87 & +27 & 21 & 12.0 & 0.620 & D &$ 1.68\pm0.23$& 24.88 & 27.5 & 27.4 & 26.79 & 26.23 & 26.54 & 1.03 &  0.25 \\ 
  59788 & 13 & 25 & 10.24 & +27 & 21 & 53.2 & 0.628 & D &$ 1.34\pm0.14$& 25.02 & 27.8 & 28.0 & 27.05 & 26.68 & 27.04 & 1.50 &  0.01 \\ 
  68251 & 13 & 23 & 40.67 & +27 & 23 & 45.6 & 0.635 & D &$ 1.44\pm0.00$& 23.46 & 25.8 & 25.6 & 25.04 & 24.73 & 24.87 & 1.10 &  0.17 \\ 
  70071 & 13 & 24 & 34.91 & +27 & 24 & 10.2 & 0.623 & D &$ 0.26\pm0.00$& 22.35 & 24.6 & 24.4 & 24.04 & 23.77 & 24.08 & 1.24 & -0.04 \\ 
  76702 & 13 & 24 &  5.75 & +27 & 25 & 37.3 & 1.186 & D &$-5.99\pm0.00$& 22.93 & 24.6 & 24.4 & 24.39 & 24.17 & 23.96 & 1.04 &  0.43 \\ 
  78892 & 13 & 24 & 15.26 & +27 & 25 & 60.0 & 0.615 & F &$-0.23\pm0.01$& 23.10 & 25.8 & 25.6 & 24.95 & 24.42 & 24.93 & 1.08 &  0.02 \\ 
  96705 & 13 & 24 & 25.59 & +27 & 29 & 47.2 & 0.632 & F &$ 2.92\pm0.67$& 25.27 & 28.5 & 27.9 & 27.89 & 27.04 & 27.04 & 1.69 &  0.85 \\ 
  99588 & 13 & 23 & 57.78 & +27 & 30 & 30.1 & 0.636 & F &$-1.82\pm0.43$& 24.21 & 26.9 & 26.8 & 26.41 & 25.99 & 26.34 & 1.42 &  0.08 \\ 
 110439 & 13 & 25 & 24.72 & +27 & 32 & 44.3 & 0.629 & D &$ 0.06\pm0.00$& 23.45 & 25.9 & 25.8 & 25.36 & 24.99 & 25.46 & 1.41 & -0.10 \\ 
 122518 & 13 & 25 & 13.45 & +27 & 35 & 17.9 & 1.174 & D &$-1.03\pm0.00$& 25.35 & 28.8 & 28.1 & 28.20 & 27.84 & 27.04 & 1.89 &  1.16 \\ 
 136295 & 13 & 25 &  5.54 & +27 & 38 & 10.0 & 0.244 & D &$-3.23\pm0.01$& 22.27 & 24.2 & 24.0 & 23.69 & 23.71 & 23.89 & 1.30 & -0.20 \\ 
 165225 & 13 & 24 & 53.38 & +27 & 43 & 57.7 & 0.637 & D &$-13.95\pm0.77$& 24.67 & 27.0 & 26.7 & 26.10 & 25.81 & 25.83 & 1.04 &  0.26 \\ 
 168136 & 13 & 24 &  3.01 & +27 & 44 & 34.7 &  ...  & F &$-3.69\pm0.33$& 23.88 & 25.5 & 25.4 & 25.30 & 25.09 & 24.79 & 1.03 &  0.51 \\ 
\hline
    \end{tabular}

\end{center}

\end{table*}

\clearpage


\begin{table*}

\tabcolsep 2pt

\begin{center}

(c) unclear objects

    \begin{tabular}{rrrrrrrccrccccccccr}
\hline
\hline
ID & \multicolumn{3}{c}{$\alpha$ (J2000.0)} &
     \multicolumn{3}{c}{$\delta$ (J2000.0)} &
     \hspace{7pt}$z$\hspace{8pt} & I & 
     \multicolumn{1}{c}{$S_W$} &
    $NB816$ & $B$ & $V$ & $R$ & $i'$ & $z'$ & $i'-NB816$ & $R-z'$ \\
\hline
  17721 & 13 & 25 &  9.51 & +27 & 13 & 29.4 &  ...  & D &$ 2.93\pm0.53$& 24.67 & 28.8 & 28.1 & 28.09 & 26.81 & 25.45 & 1.73 &  2.63 \\ 
  34503 & 13 & 25 &  7.24 & +27 & 16 & 38.7 &  ...  & D &$ 0.92\pm1.30$& 25.50 & 28.8 & 28.1 & 28.20 & 27.84 & 26.84 & 1.94 &  1.36 \\ 
  41967 & 13 & 23 & 58.35 & +27 & 18 &  1.4 &  ...  & F &$ 0.01\pm0.28$& 25.73 & 28.8 & 28.1 & 28.20 & 27.84 & 27.04 & 2.03 &  1.16 \\ 
  74901 & 13 & 23 & 58.83 & +27 & 25 &  5.1 &  ...  & F &$-2.86\pm2.14$& 25.49 & 28.8 & 28.1 & 28.20 & 27.49 & 27.04 & 1.88 &  1.16 \\ 
  89624 & 13 & 24 & 24.80 & +27 & 28 & 11.9 &  ...  & F &$ 2.04\pm0.97$& 25.70 & 28.8 & 28.1 & 28.20 & 27.84 & 27.04 & 1.84 &  1.16 \\ 
  91179 & 13 & 24 & 17.84 & +27 & 28 & 33.4 &  ...  & F &$ 2.20\pm0.92$& 25.05 & 28.8 & 28.1 & 28.20 & 27.84 & 27.04 & 2.32 &  1.16 \\ 
  97631 & 13 & 24 &  0.34 & +27 & 30 &  1.0 &  ...  & F &$ 2.01\pm1.40$& 24.98 & 28.8 & 28.1 & 28.20 & 27.16 & 26.73 & 1.87 &  1.47 \\ 
  98040 & 13 & 24 & 32.88 & +27 & 30 &  8.8 &  ...  & F &$-6.53\pm2.32$& 24.22 & 28.8 & 28.1 & 28.20 & 26.44 & 25.45 & 1.83 &  2.75 \\ 
 105644 & 13 & 23 & 49.49 & +27 & 31 & 42.7 &  ...  & F &$-0.13\pm0.80$& 24.74 & 28.8 & 28.1 & 28.20 & 27.05 & 26.04 & 1.56 &  2.16 \\ 
 111685 & 13 & 24 &  8.75 & +27 & 32 & 58.8 &  ...  & F &$-0.60\pm0.25$& 25.70 & 28.8 & 28.1 & 28.20 & 27.84 & 27.04 & 2.08 &  1.16 \\ 
\hline
    \end{tabular}

\end{center}

Notes. 

(1): ID in the NB816-detected catalog.

(4): Redshift. For LAEs 1215.67 \AA\ is adopted as 
the Ly$\alpha$ wavelength.

(5): Instrument used. D $=$ DEIMOS, F $=$ FOCAS.
The nearby object 76702 was observed with FOCAS as well.
We have adopted the DEIMOS spectrum with a better spectral resolution.

(6): Weighted skewness. 

(7) -- (12): Magnitudes in the six bandpasses. 
NB816 magnitude is the MAG$\_$AUTO magnitude, 
and the others are $2''$-aperture magnitudes. 
Magnitudes fainter than the $2\sigma$ magnitudes have been replaced 
with the $2\sigma$ magnitudes.

(13),(14): Colors measured on a $2''$ aperture.

\end{table*}


\begin{table*}
  \caption{Breakdown of the spectroscopic sample
in terms of the selection criteria for our photometric sample
    \label{tab:obj_spec_num}
  }

\begin{center}

    \begin{tabular}{cccc}
\hline
class    & \multicolumn{2}{c}{selection criteria}        & total \\
         & \multicolumn{2}{c}{------------------------}  &       \\
         & satisfy  & not satisfy                        &       \\
\hline
\hline
LAE      &   28 (27) &   6  (5) &   34 (32) \\
nearby   &    1  (1) &  18 (15) &   19 (16) \\
unclear  &   10  (8) &   0  (0) &   10  (8) \\
total    &   39 (36) &  24 (20) &   63 (56) \\
\hline
\end{tabular}

\end{center}

Notes.

Numbers in parentheses are the number of objects lying 
in the region used to construct the photometric sample 
of LAEs (See Fig. \ref{fig:xy} and section 4).

\end{table*}

\clearpage


\begin{table*}
  \caption{Ly$\alpha$ and continuum properties of confirmed LAEs
    \label{tab:LAE_prop}
  }

    \begin{tabular}{rrrcrrrrrrrcr}
\hline
\hline
     \multicolumn{1}{c}{ID}  & 
     \multicolumn{1}{c}{$z$} & 
     \multicolumn{1}{c}{FWHM}& 
     \multicolumn{1}{c}{$L({\rm Ly}\alpha)^{\rm sp}$}& 
     $L({\rm Ly}\alpha)$ & 
     \multicolumn{3}{c}{$W{\rm_{0}^{a}}(z')$} &
     \multicolumn{3}{c}{$W{\rm_{0}^{a}}(i')$} &
     $M_{\rm UV}$ & 
     SFR \\
     \multicolumn{1}{c}{(1)} & 
     \multicolumn{1}{c}{(2)} & 
     \multicolumn{1}{c}{(3)} & 
     \multicolumn{1}{c}{(4)} & 
     \multicolumn{1}{c}{(5)} & 
     \multicolumn{3}{c}{(6)} & 
     \multicolumn{3}{c}{(7)} & 
     \multicolumn{1}{c}{(8)} & 
     \multicolumn{1}{c}{(9)} \\
\hline
   18699 & 5.645 &  9.0 &    4.8 &$ 7.2\pm3.6$ & 125 &  57 & 329 &$\infty$&$\infty$&$\infty$&   ...&  ... \\ 
   20087 & 5.724 &  5.3 &   15.3 &$18.7\pm0.7$ & 135 & 105 & 184 &  92   &  67   & 139   &$-20.49$ &  8.5 \\ 
   20495 & 5.697 & 10.5 &    5.5 &$ 7.9\pm1.0$ & 112 &  66 & 268 &  86   &  40   & 780   &$-19.72$ &  4.2 \\ 
   23759 & 5.687 &  1.8 &    0.9 &$ 7.5\pm1.1$ &  39 &  29 &  57 &  76   &  37   & 364   &$-20.86$ & 11.9 \\ 
   27787 & 5.707 &  6.0 &    2.8 &$ 5.3\pm0.9$ & 210 &  95 & 701 & 160   &  47  &$\infty$&   ...   &  ... \\ 
   31858 & 5.692 &  3.8 &    7.2 &$ 3.8\pm0.9$ & 189 &  80 & 618 &3321   &  63  &$\infty$&   ...   &  ... \\ 
   31765 & 5.691 &  6.2 &    1.8 &$10.8\pm1.1$ &  76 &  55 & 121 &  53   &  33   & 104   &$-20.52$ &  8.8 \\ 
{\bf 36334} & 5.705 &  4.8 &    7.2 &$ 7.1\pm0.9$ &  22 &  19 &  26 &   9   &   7   &  12   &$-21.42$ & 20.0 \\ 
   39849 & 5.692 &  4.8 &    2.9 &$13.4\pm0.9$ & 146 &  99 & 256 &1478   & 170  &$\infty$&$-20.09$ &  5.9 \\ 
   42576 & 5.684 &  6.3 &    4.0 &$10.0\pm1.0$ & 115 &  78 & 208 & 409   & 102  &$\infty$&$-19.98$ &  5.3 \\ 
   46904 & 5.665 &  8.1 &    4.8 &$13.9\pm1.7$ &  38 &  31 &  45 &  31   &  21   &  48   &$-21.57$ & 23.1 \\ 
{\bf 48328} & 5.693 &  0.6 &    1.4 &$ 2.8\pm0.9$ &  51 &  27 & 120 &  19   &   8   &  60   &   ...   &  ... \\ 
   50215 & 5.682 &  1.5 &    2.2 &$ 5.2\pm1.0$ & 135 &  65 & 421 &$\infty$&  86 &$\infty$&   ...   &  ... \\ 
   61418 & 5.748 &  6.7 &   18.4 &$11.1\pm1.1$ & 103 &  67 & 203 & 354   &  94  &$\infty$&$-20.21$ &  6.5 \\ 
   70600 & 5.654 &  4.0 &    6.6 &$15.1\pm2.3$ & 225 & 138 & 487 &  87   &  34   & 482   &$-19.70$ &  4.1 \\ 
   73078 & 5.700 & 23.0 &   10.2 &$ 3.1\pm0.9$ & 118 &  54 & 378 &$\infty$&  85 &$\infty$&   ...   &  ... \\ 
   84305 & 5.695 &  5.8 &    3.5 &$12.0\pm1.1$ &  81 &  60 & 121 &  67   &  42   & 143   &$-20.57$ &  9.1 \\ 
   84720 & 5.682 & 10.9 &    2.7 &$10.3\pm1.4$ &  77 &  48 & 145 &  69   &  32   & 289   &$-20.47$ &  8.4 \\ 
   93966 & 5.721 &  0.0 &    3.3 &$ 6.2\pm0.8$ & 164 &  92 & 475 &$\infty$& 183 &$\infty$&   ...   &  ... \\ 
{\bf 95588} & 5.738 &  6.8 &    1.5 &$ 5.9\pm0.9$ &  57 &  40 & 100 &  24   &  16   &  37   &$-20.14$ &  6.2 \\ 
{\bf 96007} & 5.696 &  6.1 &    1.2 &$ 4.8\pm1.1$ &  49 &  29 & 108 &  24   &  11   &  61   &$-20.08$ &  5.8 \\ 
  108164 & 5.672 &  7.3 &    1.3 &$ 3.3\pm1.3$ & 167 &  68 & 635 & 109   &  21  &$\infty$&   ...   &  ... \\ 
  113271 & 5.710 &  9.5 &    7.0 &$25.4\pm0.8$ &  39 &  36 &  42 &  57   &  49   &  67   &$-22.17$ & 40.1 \\ 
  121315 & 5.716 &  0.0 &    3.0 &$ 4.9\pm0.8$ & 148 &  71 & 505 &  52   &  26   & 276   &   ...   &  ... \\ 
{\bf 124530} & 5.689 &  2.2 &    2.7 &$ 9.4\pm1.1$ & 131 &  78 & 307 &  31   &  20   &  61   &$-19.65$ &  3.9 \\ 
  124783 & 5.720 & 12.0 &    4.0 &$ 5.2\pm0.8$ &  35 &  26 &  48 &  35   &  22   &  65   &$-20.58$ &  9.2 \\ 
  126848 & 5.718 & 18.8 &    4.9 &$ 5.0\pm0.8$ &  91 &  53 & 221 &  73   &  33   & 592   &   ...   &  ... \\ 
  132343 & 5.697 &  3.6 &    3.4 &$ 8.1\pm1.0$ & 144 &  77 & 387 &  98   &  43  &$\infty$&   ...   &  ... \\ 
  152586 & 5.681 &  7.6 &    2.6 &$13.9\pm1.3$ &  90 &  63 & 146 &  82   &  45   & 197   &$-20.59$ &  9.3 \\ 
  154296 & 5.715 &  4.9 &    0.8 &$ 4.0\pm0.9$ & 131 &  60 & 431 & 133   &  35  &$\infty$&   ...   &  ... \\ 
  158036 & 5.676 & 11.2 &    1.9 &$ 8.0\pm1.4$ &  92 &  54 & 203 & 211   &  52  &$\infty$&$-20.00$ &  5.4 \\ 
  163079 & 5.676 &  4.8 &    1.8 &$10.6\pm1.4$ &  96 &  64 & 178 &  66   &  35   & 193   &$-20.24$ &  6.8 \\ 
{\bf 166310} & 5.698 &  9.3 &    3.4 &$12.6\pm0.9$ &  21 &  19 &  24 &   8   &   6   &   9   &$-22.08$ & 36.7 \\ 
  168127 & 5.674 &  5.5 &    3.9 &$15.4\pm1.4$ & 261 & 143 & 656 & 146   &  69   &1423   &   ...   &  ... \\ 
\hline
\end{tabular}

Notes. 

(1): ID in the NB816-detected catalog.
Shown in bold font are 
objects not satisfying the photometric selection criteria.

(2): redshift.

(3): FWHM of the Ly$\alpha$ line in the observed frame (\AA).

(4): Ly$\alpha$ luminosity ($\times 10^{42}$ erg s$^{-1}$) 
calculated from the spectrum. 

(5): Ly$\alpha$ luminosity ($\times 10^{42}$ erg s$^{-1}$) 
calculated from the NB816- and $z'$-band photometry combined 
with the spectroscopic redshift.

(6): $W{\rm_{0}^{a}}(z')$ (\AA). 
The first, second, and third columns 
are the central value ($W_c$), $1\sigma$ lower limit ($W_{-}$), 
and $1\sigma$ upper limit ($W_{+}$).
These values are defined as 
$\int_{0}^{W_{c}} P(W)dW = 0.5$ and 
$\int_{W_{-}}^{W_{c}} P(W)dW = \int_{W_{c}}^{W_{+}} P(W)dW = 0.34$, 
where $P(W)dW$ is the probability distribution of EW 
derived from Monte Carlo simulations.

(7): $W{\rm_{0}^{a}}(i')$ (\AA). 
The first, second, and third columns 
are the central value, $1\sigma$ lower limit, and 
$1\sigma$ upper limit.
The definition of these values is the same as 
for $W{\rm_{0}^{a}}(z')$.

(8): Far-UV continuum absolute magnitude (AB).

(9): Star formation rate calculated from $M_{\rm UV}$ 
($M_\odot$ yr$^{-1}$).

\end{table*}

\end{document}